\documentclass{aa}
\usepackage{txfonts}
\usepackage{graphicx}
\usepackage{aalongtable}

\newcommand{\ee}{\end{equation}}
\newcommand{\be}{\begin{equation}}
\newcommand{\ec}{\end{center}}
\newcommand{\bc}{\begin{center}}
\newcommand{\eea}{\end{eqnarray}}
\newcommand{\bea}{\begin{eqnarray}}
\newcommand{\bd}{\begin{description}}
\newcommand{\ed}{\end{description}}
\newcommand{\bi}{\begin{itemize}}
\newcommand{\ei}{\end{itemize}}

\newcommand{\mini}{\mbox{$M_{\rm i}$}}
\newcommand{\diff}{\mbox{${\rm d}$}}

\newcommand{\bvo}{\mbox{$(B\!-\!V)_{0}$}}

\newcommand{\bv}{\mbox{$(B\!-\!V)$}}

\newcommand{\vk}{\mbox{$(V\!-\!K)$}}
\newcommand{\ks}{\mbox{$K_{\mathrm s}$}}
\newcommand{\jh}{\mbox{$J\!-\!H$}}
\newcommand{\jk}{\mbox{$J\!-\!K$}}

\newcommand{\mv}{\mbox{$M_{V}$}}

\newcommand{\ebv}{\mbox{$E_{B\!-\!V}$}}
\newcommand{\feh}{\mbox{\rm [{\rm Fe}/{\rm H}]}}

\newcommand{\Msun}{\mbox{$M_{\odot}$}}
\newcommand{\Teff}{\mbox{$T_{\rm eff}$}}

\newcommand{\logg}{\mbox{$\log g$}}
\newcommand{\etal}{\mbox{\rm et~al.}}

\newcommand{\comment}[1]{}
\newcommand{\beq}{\begin{equation}}
\newcommand{\eeq}{\end{equation}}
\newcommand{\beqa}{\begin{eqnarray}}
\newcommand{\eeqa}{\end{eqnarray}}
\newcommand{\bali}{\begin{alignat}}
\newcommand{\eali}{\end{alignat}}
\newcommand{\benu}{\begin{enumerate}}
\newcommand{\eenu}{\end{enumerate}}
\newcommand{\bite}{\begin{itemize}}
\newcommand{\eite}{\end{itemize}}
\newcommand{\bdes}{\begin{description}}
\newcommand{\edes}{\end{description}}

\begin{document}

\title{Basic physical parameters of a selected sample of evolved stars
\thanks {Based on observations collected at the ESO - La Silla, partially under
the ON-ESO agreement for the joint operation of the 1.52\,m ESO telescope}
}
\subtitle{}
\author{L. da Silva\inst{1} \and L. Girardi\inst{2} \and L. Pasquini\inst{3} \and
J. Setiawan \inst{4} \and O. von der L\"uhe\inst{5} \and J.R. de Medeiros\inst{6} \and A. Hatzes\inst{7} \and
M. P. D\"ollinger\inst{3} \and  A. Weiss\inst{8}
}
\offprints{licio@on.br}
\institute{Observat\'orio Nacional - MCT, Rio da Janeiro, Brazil
\and
Osservatorio Astronomico di Padova -- INAF, Padova, Italy
\and
European Southern Observatory, Garching bei M\"unchen, Germany
\and
Max-Planck-Institut f\"ur Astronomie, Heidelberg, Germany
\and
Kiepenheuer-Institut f\"ur Sonnenphysik, Freiburg, Germany
\and
Departamento de F\'isica -- UFRN, Natal, Brazil
\and
Th\"uringer Landessternwarte, Tautenburg, Germany
\and
Max-Planck-Institut f\"ur Astrophysik, Garching bei M\"unchen, Germany
}
\date{received 28 Febuary 2006/ Accepted 28 Febuary 2006}
\abstract{
We present the detailed spectroscopic analysis of 72 evolved stars,
which were previously studied for accurate radial velocity variations.
Using one Hyades giant and another well studied star as the reference 
abundance, we determine the [Fe/H] for the whole sample.
These metallicities, together with the \Teff\ values and the absolute
$V$-band magnitude derived from Hipparcos parallaxes, are used to estimate
basic stellar parameters (ages, masses, radii, \bvo\ and \logg)
using theoretical isochrones and a 
Bayesian estimation method.
The \bvo\ values so estimated turn out to be in excellent agreement
(to within $\sim0.05$~mag) with the observed \bv, confirming the
reliability of the \Teff--\bvo\ relation used in the isochrones.
On the other hand, the estimated \logg\ values are typically
0.2~dex lower than those derived from spectroscopy;
this effect has a negligible impact on \feh\ determinations.
The estimated diameters $\theta$ have been compared with 
limb darkening-corrected ones
measured with independent methods, finding an agreement better
than 0.3~mas within the $1<\theta<10$~mas interval (or, alternatively,
finding mean differences of just $6$~\%).
We derive the age-metallicity relation for the solar neighborhood;
for the first time to our knowledge, such a relation has been
derived from observations of field giants rather than from
open clusters and field dwarfs and subdwarfs.
The age-metallicity relation is characterized by close-to-solar
metallicities for stars younger than $\sim4$~Gyr, and by a large
\feh\ spread with a trend towards lower metallicities for higher ages.
In disagreement with other studies, we find that the \feh\ dispersion of
young stars (less than 1~Gyr) is comparable to the observational errors,
indicating that stars in the solar neighbourhood are formed from
interstellar matter of quite homogeneous chemical composition.
The three giants of our sample which have been proposed to host planets
are not metal rich; this result is at odds with those for main sequence stars.
However, two of these stars have masses much larger than a solar mass so
we may be sampling a different stellar population from most radial velocity
searches for extrasolar planets.
We also confirm the previous indication that the radial velocity
variability tends to increase along the RGB, and in particular
with the stellar radius.
}

\authorrunning{L. da Silva  et al.}
\titlerunning{Basic physical parameters of Giant stars}
\maketitle
%
\section{Introduction}

It has recently become evident from high accuracy radial velocity (RV) measurements
that late-type (G and K) giant stars are RV variables
(see e.g. Hatzes \& Cochran 1993, 1994; Setiawan et al. 2003, 2004).
These variations occur on two greatly different timescales:
short term variability with periods in the range 2--10 days, and long term
variations with periods greater than several hundreds of days. The short
term variations are due to stellar oscillations. The cause
of the long term variability is not clear and at least three mechanisms,
i.e., low mass companions, pulsations and surface activity, have been proposed.
The first results of our long term study of the nature of the variability of K giants
have shown that all three mechanisms are likely contributors to long term RV variability,
although their dependence on the star's fundamental characteristics remain unknown
(Setiawan et al. 2003, 2004).

The purpose of this paper is to accurately determine the radii, temperatures,
masses, and chemical composition of the stars of our sample, in order to understand
better how RV variability and stellar characteristics are related.
We are particularly interested in determining to what extent the stars found to host
planetary system class bodies (Setiawan et al. 2003a, 2005) exhibit high metallicity,
as has been found for the dwarfs hosting giant exoplanets ( Gonzales 1997; Santos et al. 2004).

The present sample is composed of the stars published by Setiawan et al. (2004).  
The selection criteria and the observations were explained there.
We recall that observations were obtained with the FEROS spectrograph at the ESO
1.5m telescope (Kaufer et al. 1999), with a resolving power of 50000 and a
signal-to-noise ratio exceeding 150 in the red part of the spectra.

\section{Atmospheric parameters determination}
\label{chemical}

The spectroscopic analysis has been made in LTE, using a modified version of
Spite's (1967) code.
MARCS plane parallel atmosphere models are used.
Gustafsson et al. (1975) models were used for the giants, while
Edvardsson et al. (1993) models were used for the few dwarfs of our sample.
No major spurious effects are expected by using those two sets of model atmospheres
(Pasquini et al. 2004).
Equivalent widths of spectral lines were measured using the DAOSPEC package
(Pancino \& Stetson 2005).
The spectra of several stars were cross checked by measuring the equivalent widths with MIDAS,
resulting in a very good agreement of the two methods.
The line list and corresponding atomic data were those adopted by Pasquini et al. (2004).

Atmospheric parameters (effective temperatures, surface gravities, microturbulence velocities,
and metallicities) have been obtained using an iterative and self-consistent procedure.
The atmospheric parameters were determined from the spectroscopic data in the conventional way.
The effective temperature, \Teff, is determined by imposing that the Fe I abundance does
not depend on the excitation potentials of the lines.
The microturbulence velocity is determined by imposing that the Fe I abundance is independent
of the line equivalent widths.
The Fe I/Fe II ionization equilibrium has been used to determine the surface gravity $g$.
The method used here is described in detail by \cite{peloso05}, with the difference that temperature and gravity are also free parameters in the present work.
The initial values for \Teff\ and \logg\ to start the process were derived in two steps.
The first step involves the \bv\ photometry and parallax from Hipparcos, assuming for all stars
solar metallicity and a 1.5~\Msun\ mass, and adopting the \cite{alonso99} scale.
The initial microturbulence velocity was 1.0 km\,s$^{-1}$.
In the second step, the values for \logg\ were obtained from the Girardi et al. (2000)
evolutionary tracks by interpolation, using the previously determined values for \Teff,
abundance and \mv\ .

\subsection{Effective temperature}

The \vk\ index, which  is used by many authors, is probably the best photometric
\Teff\ indicator for G-K giants (Plez et al. 1992, Ramirez \& Mel\'endez 2004).
The only source of $K$-band photometry for our sample is the 2MASS Catalog \cite{2mass},
but its authors caution that stars with $\ks<3$ are saturated, and the error of their colors
is larger.
Many stars of our sample belong to this class and the rest is not much fainter.
We compared the \Teff\ values obtained from \vk\ with those from \bv\ to verify whether we
can use the 2MASS catalog colors to determine the effective temperature of our sample.
For both indices we used the \cite{alonso99} calibrations for the stars with $\logg<3.5$
and the \cite{alonso96} calibration for the remaining dwarfs.
The 2MASS \ks\ magnitude was converted to the CTS system (\cite{alonso98}) via the
CIT system \cite{2mass}.
The \vk\ indices of the stars with $\logg>3.5$ were converted to the Johnson system using
the relation given by \cite{alonso98}.
The [Fe/H] values needed to apply those relations were taken from Table~\ref{tab:01}.

\begin{figure}
  \centering
  \resizebox{6.5cm}{!}{\includegraphics{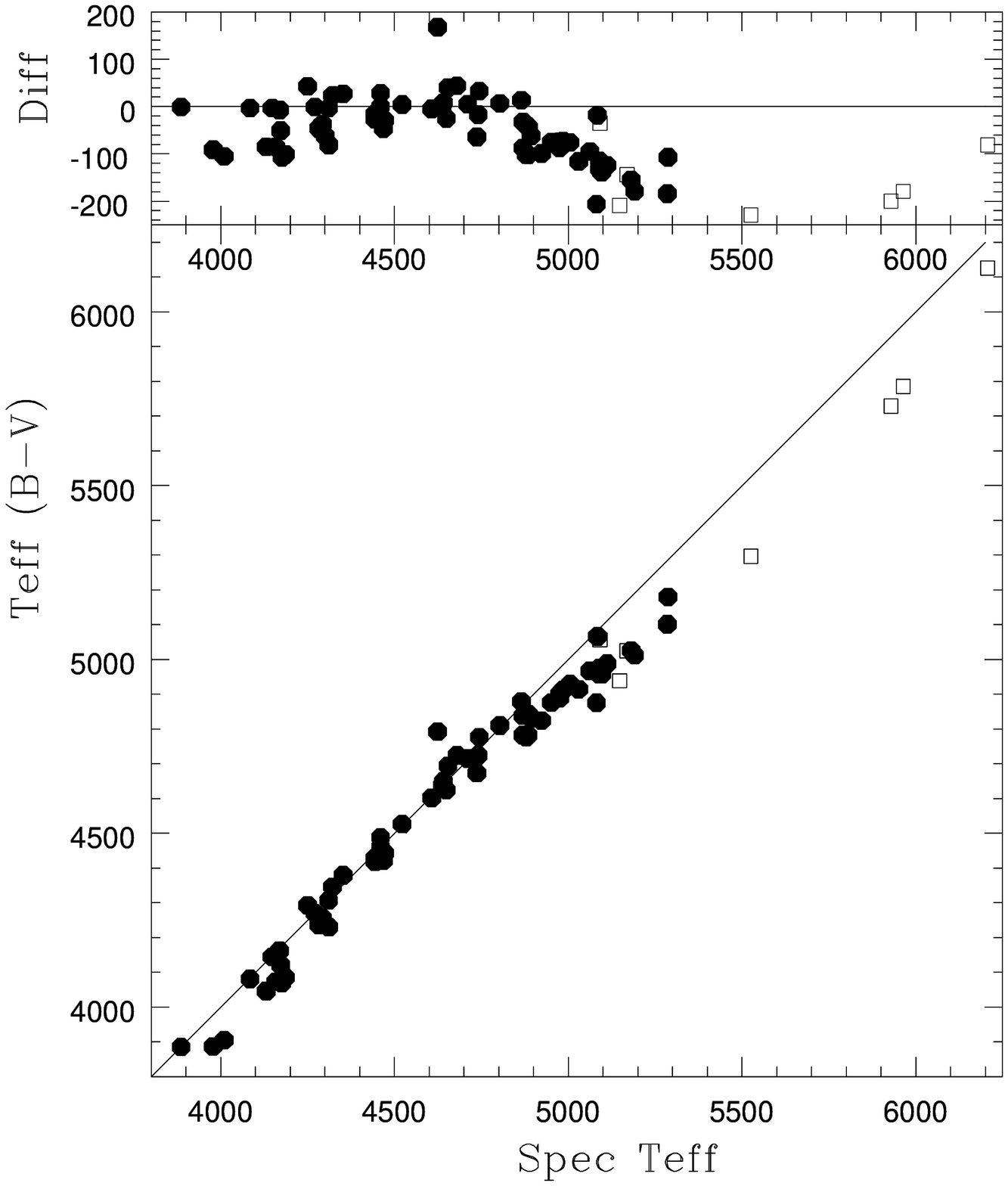} }
  \resizebox{6.5cm}{!}{\includegraphics{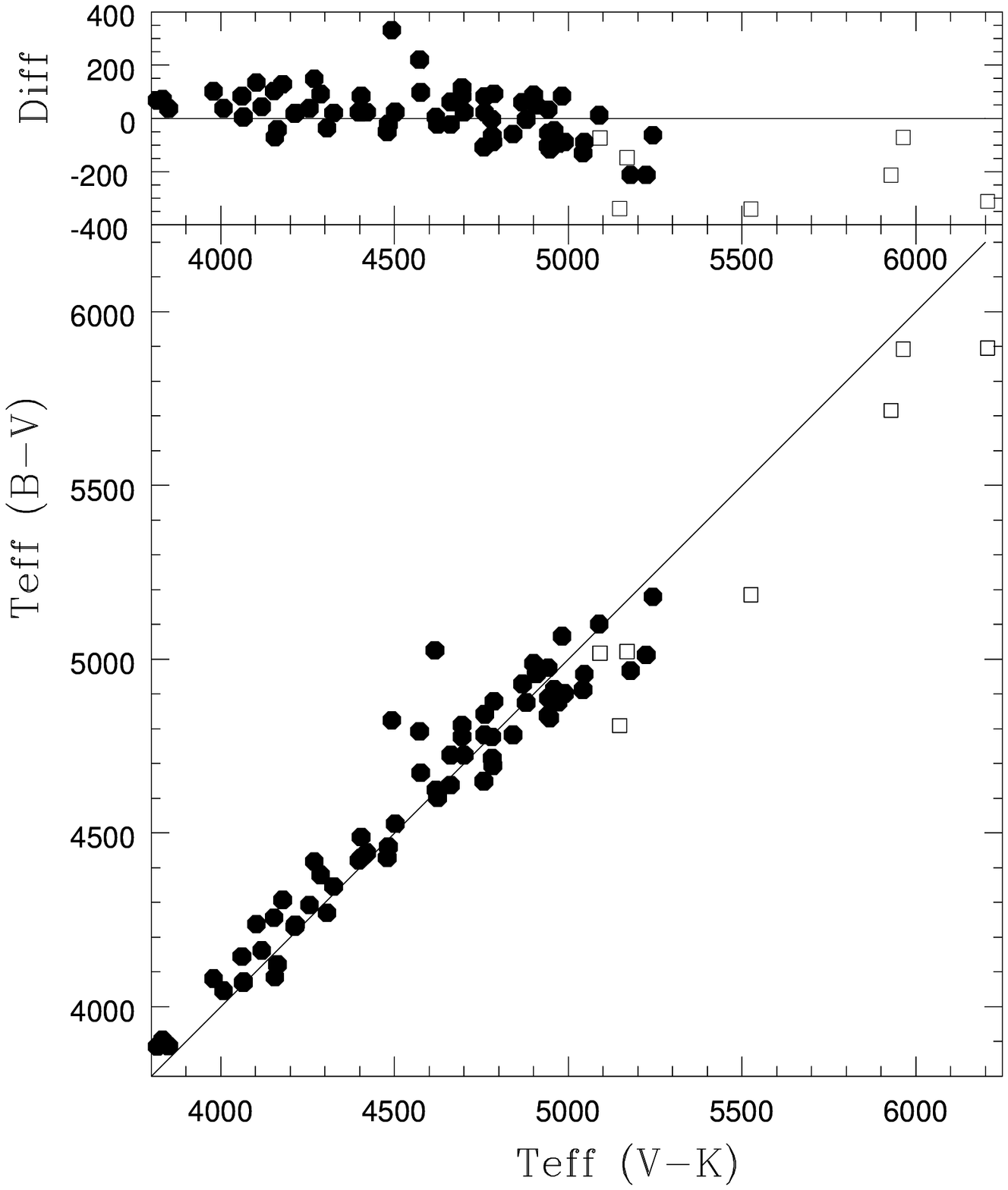} }
  \caption{\Teff\ values obtained from \bv\ compared with the values obtained from the
    spectroscopic analyses (left panel), and compared with \Teff\ obtained from \vk\
    (right panel).
    Giants are represented by circles and dwarfs by squares in both panels.
The upper panels show the differences between the  values
  }
  \label{temp}
\end{figure}

The results are presented Fig.~\ref{temp}.
It is evident that the $\Teff_{(V\!-\!K)}$ vs. $\Teff_{(B\!-\!V)}$  relation has a much larger dispersion than $\Teff_{\rm spec}$ vs. $\Teff_{(B\!-\!V)}$.
The dispersion of the relation using \jh\ and \jk\ is greater.
These findings confirm that the colors of 2MASS are unsuitable to determine the \Teff\ of
bright stars, in particular of our sample.
On the other hand, we see in Fig.~\ref{temp} that $\Teff_{\rm spec}$ is in good agreement
with $\Teff_{(B\!-\!V)}$ for stars with $4200 \le \Teff \le 5200$, while there is less
agreement for cooler and hotter stars.
This result is expected for cold stars -- because is known that \bv\ is not a good
\Teff\ indicator for them -- but it is unexpected for those hotter stars.
Noting that the hotter stars of our sample are dwarfs, we compared the \Teff\ which we
determined from spectroscopy and from the \bv\ index with the values given in
\cite{peloso05} for a sample of G-K dwarfs.
The temperatures of the \cite{peloso05} stars were carefully determined using criteria
different from the Fe I excitation equilibrium.
The agreement between their and our results is better for our spectroscopy values than for
the \bv\ values. Note that other authors also found differences between spectroscopic and photometric  \Teff\ for dwarfs of the same order or larger than the differences we found (e.g. Reddy et al. 2003; Santos et al. 2004; Ramirez \& Melendez 2004; Luck \& Heiter 2005). Moreover, for $\Teff \lesssim 4000 K$ there is a saturation of \bv\ and this colour index is no longer useful (see Alonso et al. 1999 for a discussion about this point). For these reasons, we will adopt our spectroscopic temperatures in the followin analysis.
				   
\subsection{Microturbulence and iron abundance}

The determination of microturbulence in giants may be difficult, as discussed in detail in previous papers (e.g. McWilliam 1990).
In order to test the sensitivity of our analysis with regard to this parameter,
we have performed the process described above by starting with two very different
values of microturbulence (e.g. 1 and 2.5 km\,s$^{-1}$) and letting the system converge freely.  
We found for almost all stars a very good convergence, resulting in the same
final values for all parameters, including microturbulence.
Only for the four coolest stars of the sample does the analysis converge to microturbulence
velocities that differ by up to 0.2 km\,s$^{-1}$, and to Fe abundances that differ by
up to 0.1 dex.
These results were obtained using the same line lists and equivalent widths.
This result indicates that the solution space may be degenerate for the coolest stars,
and that fairly large errors may intrinsically be present.
We also notice that for these stars, we may be exceeding the validity limit of the grid of
our atmosphere models.
The atmospheric parameters (effective temperature, gravity, metallicity and microturbulence)
which we determined from the spectral analysis are presented in Table~\ref{tab:01}.  In a coming paper (da Silva et al., in preparation) we will present and analyze the abundances of some  other elements and we will publish the equivalent widths and atomic data (log gf and excitation potential) of all the lines used in our analysis, for all sample stars.

\begin{table*}
\caption{
  Retrieved atmospheric parameters for the entire star sample.
  Note that metallicities are 0.07 dex larger in comparison to the scale
  derived independently from the analysis of HD~27371 and HD~113226; see text.
  Column 1: HD number; Column 2: Spectroscopic effective temperature;  
  Column 3: Iron content (normalized to the Sun); Column 4: Spectroscopic gravity;
  Column 5: Microturbulence (km\,s$^{-1}$).
}
\begin{tabular}{lrrrr|lrrrr|lrrrr}
\hline\hline
HD   & \Teff & Fe1 & \logg & $\xi$ & HD & \Teff & Fe1  & \logg & $\xi$ & HD & \Teff & Fe1  & \logg & $\xi$ \\
(1) & (2) & (3) & (4) & (5) & (1) & (2) & (3) & (4) & (5) & (1) & (2) & (3) & (4) & (5) \\
\hline
2114 & 5288 & 0.04 & 2.9 & 1.8 &    61935 & 4879 & 0.06 & 3.0 & 1.6 &       125560 & 4472 & 0.23 & 2.4 & 1.6\\
2151 & 5964 & 0.04 & 4.3 & 1.0 &    62644 & 5526 & 0.19 & 4.1 & 1.0&       131109 & 4158 & 0.00 & 1.9 & 1.5\\
7672 & 5096 & $-$0.26 & 3.3 & 1.5 &    62902 & 4311 & 0.40 & 2.5 & 1.4 &       136014 & 4869 & $-$0.39 & 2.7 & 1.5\\
11977 & 4975 & $-$0.14 & 2.9 & 1.6 &   63697 & 4322 & 0.20 & 2.2 & 1.4 &       148760 & 4654 & 0.20 & 3.0 & 1.4\\
12438 & 4975 & $-$0.54 & 2.5 & 1.5 &    65695 & 4468 & $-$0.07 & 2.3 & 1.5 &       151249 & 3886 & $-$0.30 & 1.0 & 1.3\\
16417 & 5936 & 0.26 & 4.3 & 1.2 &    70982 & 5089 & 0.04 & 3.0 & 1.6 &        152334 & 4169 & 0.13 & 2.0 & 1.4\\
18322 & 4637 & 0.00 & 2.7 & 1.4 &    72650 & 4310 & 0.13 & 2.1 & 1.5 &       152980 & 4176 & 0.08 & 1.8 & 1.6\\
18885 & 4737 & 0.17 & 2.7 & 1.5 &    76376 & 4282 & $-$0.03 & 1.9 & 1.6 &      159194 & 4444 & 0.21 & 2.3 & 1.6\\
18907 & 5091 & $-$0.54 & 3.8 & 0.9 &    81797 & 4186 & 0.07 & 1.7 & 1.7 &       165760 & 5005 & 0.09 & 2.9 & 1.6\\
21120 & 5180 & $-$0.05 & 2.8 & 1.8 &    83441 & 4649 & 0.17 & 2.8 & 1.4 &       169370 & 4460 & $-$0.10 & 2.3 & 1.3\\
22663 & 4624 & 0.18 & 2.7 & 1.2 &    85035 & 4680 & 0.19 & 3.2 & 1.5 &        174295 & 4893 & $-$0.17 & 2.8 & 1.5\\
23319 & 4522 & 0.31 & 2.5 & 1.5 &    90957 & 4172 & 0.12 & 1.9 & 1.5 &       175751 & 4710 & 0.08 & 2.7 & 1.6\\
23940 & 4884 & $-$0.28 & 2.8 & 1.5 &    92588 & 5136 & 0.14 & 3.8 & 0.9 &      177389 & 5131 & 0.09 & 3.7 & 1.1\\
26923 & 6207 & 0.01 & 4.5 & 2.5 &    93257 & 4607 & 0.20 & 2.8 & 1.5 &        179799 & 4865 & 0.10 & 3.4 & 1.1\\
27256 & 5196 & 0.14 & 3.0 & 1.7 &    93773 & 4985 & 0.00 & 3.0 & 1.5 &       187195 & 4444 & 0.20 & 2.6 & 1.4\\
27371 & 5030 & 0.20 & 3.0 & 1.7 &    99167 & 4010 & $-$0.29 & 1.3 & 1.4 &      189319 & 3978 & $-$0.22 & 1.2 & 1.7\\
27697 & 4951 & 0.13 & 2.8 & 1.7 &    101321 & 4803 & $-$0.07 & 3.1 & 1.2 &      190608 & 4741 & 0.12 & 3.1 & 1.4\\
32887 & 4131 & $-$0.02 & 1.8 & 1.5 &    107446 & 4148 & $-$0.03 & 1.8 & 1.5 &      198232 & 4923 & 0.10 & 2.8 & 1.5\\  
34642 & 4870 & 0.03 & 3.3 & 1.3 &    110014 & 4445 & 0.26 & 2.2 & 1.7 &        198431 & 4641 & $-$0.05 & 2.8 & 1.2\\
36189 & 5081 & 0.05 & 2.8 & 1.9 &    111884 & 4271 & 0.01 & 2.2 & 1.5 &        199665 & 5089 & 0.12 & 3.3 & 1.4\\  
36848 & 4460 & 0.28 & 2.7 & 1.5 &    113226 & 5086 & 0.16 & 2.9 & 1.7 &        217428 & 5285 & 0.10 & 3.1 & 1.8\\  
47205 & 4744 & 0.25 & 3.2 & 1.3 &    115478 & 4250 & 0.10 & 2.1 & 1.4 &        218527 & 5084 & 0.10 & 3.1 & 1.5\\  
47536 & 4352 & $-$0.61 & 2.1 & 1.4 &    122430 & 4300 & 0.02 & 2.0 & 1.5 &        219615 & 4885 & $-$0.44 & 2.6 & 1.4\\
50778 & 4084 & $-$0.22 & 1.7 & 1.5 &    124882 & 4293 & $-$0.17 & 2.1 & 1.5 &       224533 & 5062 & 0.07 & 3.1 & 1.5\\
\end{tabular}
\label{tab:01}
\end{table*}

We still have to determine the zero point of the metallicity scale.
Pasquini et al. (2004) used a normalization to the solar spectrum, because their
sample contains many solar-type stars.
Since our sample is mainly composed of evolved stars, we prefer to fix the abundance scale
by using giants that have been well studied in the literature.
Two stars of our sample are particularly well studied, with several high quality entries
in the Cayrel de Strobel et al. (2001) catalogue,namely HD~113226 ($\epsilon$ Vir) and HD~27371
($\gamma$ Tau), a Hyades giant (the literature data for another Hyades giant in our sample,
HD~27697, are much less constrained).

Averaging all entries of the Cayrel de Strobel et al. (2001) catalogue for HD~113226,
we obtain $\feh=0.11$ and $\Teff=5048$~K, while for HD~27371 we obtain $\feh=0.11$ and
$\Teff=4967$~K (the two values of Komaro in the catalogue for HD~27371 were not considered).
Our results from Table~\ref{tab:01} are $\feh=0.16$ and $\Teff=5086$~K,
and $\feh=0.20$ and $\Teff=5030$~K for the two stars.
We conclude that it is very likely that the metallicity scale of Table~\ref{tab:01} is too high by
0.07~dex in \feh, and we will apply this correction to all our data in the following.
This correction should therefore be applied to all the entries of Table~\ref{tab:01} to derive the
correct metallicity.
Table~\ref{tabelao} shows the corrected values.
We note also that our spectroscopic temperatures for HD~113226 and HD~27371 are slightly higher
(about 50~K) than the average values reported in the literature.
Both results are in very good agreement with the results of Pasquini et al. (2004),
who found $\feh=0.06$ for the analysis of the UVES solar spectrum, and who derived
systematic differences between the spectroscopic and photometric temperatures of IC~4651 stars.
Note also that the largest differences between the average and individual \feh\ values in
the catalogue are 0.10~dex for HD~113226 and 0.09~dex for HD~27371, both of which
are larger than the discrepancy with the present analysis.

We notice also that another star in our sample has more than two entries in the
Cayrel de Strobel et al. (2001) catalogue: the moderately metal poor star HD~18907,
which is at the low metalliciy end of our sample.
The average literature value for this star is $\feh=-0.69$, while our zero-point corrected
value is $\feh=-0.54$.
Although we do not expect a strong dependence of the zero point on the metallicity, if we were to include this star in our set of calibrators,
we should apply a correction of 0.10~dex instead of 0.07.
In our analysis, HD~18907 has a very low microturbulence (0.9~km\,s$^{-1}$),
which is lower than for any other star, providing an explanation of the discrepancy with the literature. Although a second analysis of this star did not reveal any suspicious effect, we preferred not to use it to determine our zero-point. 

We could redraw Fig. 1 with the  $\Teff_{(B\!-\!V)}$ and  $\Teff_{(V\!-\!K)}$ calculated using the corrected values of [Fe/H], given in Table~\ref{tabelao}, but the $\Teff_{(B\!-\!V)}$  calibration is not very sensitive to the metallicity, and  $\Teff_{(V\!-\!K)}$ is even less so. For the stars cooler then 4500~K there is no difference at all between the \Teff\ found from the two [Fe/H] values. The largest difference in our sample is 31~K, found for the hottest star HD 26923, from $\Teff_{(B\!-\!V)}$, \Teff\ being lower for the correct (lower) metallicity. 

\begin{figure*}[ht]
\begin{center}
\resizebox{\hsize}{!}{\includegraphics{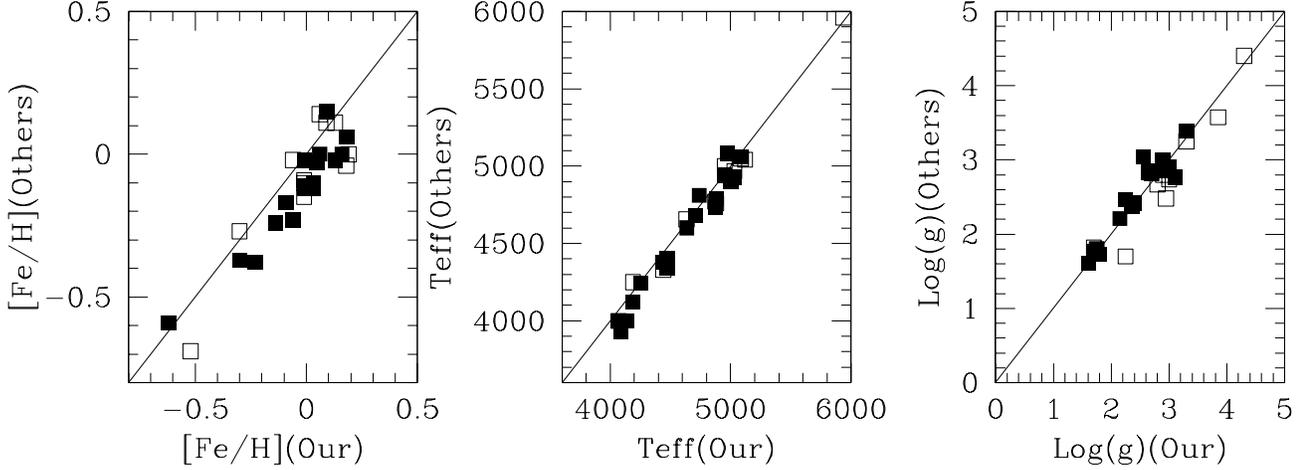}}
\end{center}
\caption{
  Comparison of our results (abscissa) and literature values (ordinate) for
  \feh, \logg\ and \Teff\ as retrieved from Cayrel de Strobel et al. (2001).  
  Open squares represent stars with more than one determination in the literature;
  filled squares represent values from McWilliam (1990).
}
\label{comparison}
\end{figure*}

We compare the \feh\ and \Teff\ values of our analysis with those from the Cayrel de Strobel
et al. (2001) catalogue in Fig.~\ref{comparison}.
Stars with more than one entry in the catalogue are shown as empty squares.  
Most of the entries are from the work by McWilliam (1990), and stars with only that
entry are shown with filled squares.
The results of the comparison are given in Table~\ref{tab:02}, and they show that the
agreement with most data in the literature is quite good, and that we tend to systematically
retrieve higher abundances (by 0.1~dex) than McWilliam (1990).
We somewhat expected such a result, because the method chosen by McWilliam to determine
microturbulence tends to result in rather high values of $\xi$.
However, his estimate of \feh\ for the Hyades giant HD~27371 is $\feh=-0.02$, which is
much lower than our value, and lower than what is considered the best estimated
abundance for this cluster.
We would therefore expect a discrepancy between our and the McWilliam results of about 0.13~dex,
which is very close to our results.
Note also that the values in Table~\ref{tab:02} are not independent, since our average computations
from the literature include the McWilliam results.

\begin{table}
\caption{
  Results of the comparison of our stellar parameters with those of the
  Cayrel de Strobel et al. (2001) catalogue.
  All differences represent our values minus the literature values.
  The third and fourth rows present mean and standard deviation of the difference with
  respect to McWilliam (1990) only.}
\begin{tabular}{lccc}
\hline
\hline
              & \feh\ & \Teff(K) & \logg\ \\
\hline
$\Delta$      & 0.07   & 39   & 0.13       \\
$\sigma$      & 0.1    & 66   & 0.23       \\
$\Delta$ McW. & 0.1    & 59   & $-$0.05      \\
$\sigma$ McW. & 0.06   & 70   & 0.16       \\
\end{tabular}
\label{tab:02}
\end{table}

\subsection{Iron abundance error estimate}

It is difficult to provide realistic error estimates for abundance and stellar parameter
determinations, such as \Teff\ and gravity.
We intend to be very careful in estimating these errors, because it is our goal to use
the parameters to derive stellar masses and radii from evolutionary tracks.  
Any error in the fundamental (spectroscopic) parameters would make the determination
of the derived stellar parameters more uncertain.

The direct comparison of our results with those of other authors is shown in Table~\ref{tab:02}.
Our Fe content is on average systematically too high by 0.07~dex,
the temperature is too high by about 40~K and gravity is too high by about 0.13~dex.
A systematic error of such a magnitude would not influence our measurements or conclusions
in a significant way.
However, the scatter about the mean discrepancies is in general more pronounced, and we
believe that the scatter represents a more realistic, albeit somewhat pessimistic,  
estimate of the uncertainty of our retrieved parameters.  
\Teff\ shows a scatter of 70~K.
The scatter of \logg\ is up to 0.2~dex and quite large.
The scatter of [Fe/H] is 0.1~dex.

An estimate of the internal error in our \feh\ determination, which is produced mainly by the uncertainties in the equivalent width measured and in the log gf is  more important
than the external error of $\sim0.1$~dex (see next section).
To estimate our internal errors in the determined metallicities,
we examined two giants of our sample in more details, one among the coolest and the other among the hottest stars, namely HD~111884 and HD~36189.
We changed each of the input parameters (\Teff , \logg\ and $\xi$)  of our 
spectroscopic analysis in turn, using the values found above for the 
scatters (70~K, 0.2~dex and 0.1~dex).
We proceeded in two ways: first, the other parameters were kept fixed; second, 
{\it which is more in agreement with our method}, the others parameters were left to adjust 
freely. The resulting changes of \feh\ were monitored and Table~\ref{tab:03} presents 
the results.  Note that step one is useful as a test because, in the method employed, when one  parameter is changed, the others also change. An error in one parameter will be reflected in the other parameters.
 Usually the errors on the  equivalent width measurements will produce greater dispersions in the 
diagrams using them as a criterion, causing uncertainties in these quantities. The continuum 
placement, for instance, will be slightly higher in some regions and slightly lower in others. To illustrate  the influence 
of the equivalent width errors on our results,  but stressing that this is not a real situation, we tested what happens when all those quantities 
are too large or too small by a constant factor. In our example we use 3\%, the results are shown in 
Table~\ref{tab:04}. The errors are not symmetric and they are much enlarged for the 
hotter than  the cooler stars. Fortunately, the continuum placement for the hotter stars are much 
more precise and a error of 3\% is unlikely.
Given these results, we will assume an internal error of 0.05~dex in our \feh\
determination.


\begin{table}
\caption{
  Sensitivity of \feh\ to changes of other stellar parameters for two stars at the
  cool and hot ends of our sample.}
\begin{tabular}{lcc}
\hline \hline
                                                                                & HD~111884                & HD~36189                \\        \hline
\multicolumn{3}{c}{Original parameters}                                                                        \\        \hline
\Teff [K]                                                                & 4271                        & 5106                        \\
\feh [dex]                                                                & 0.01                        & 0.06                        \\
\logg [dex]                                                                & 2.2                        & 2.9                        \\
$\xi$ [$km s^{-1}$]                                                & 1.5                        & 1.5                        \\        \hline
\multicolumn{3}{c}{Change of \feh, others fixed [dex]}                                        \\        \hline
$\Delta$\Teff\ = $\pm$70~K                                & $\pm$0.02                & $\pm$0.07                \\
$\Delta$\logg\ = $\pm$0.2~dex                        & $\pm$0.07                & $\pm$0.01                \\
$\Delta \xi$ = $\pm$0.2~km\,s$^{-1}$        & $\pm$0.12                & $\pm$0.06                \\
mean error, others fixed                                & $\pm$0.07                & $\pm$0.05                \\        \hline
\multicolumn{3}{c}{Change of \feh, others free [dex]}                                        \\        \hline
largest change                                                        & $\pm$0.05                & $\pm$0.04                \\        
\hline \hline
\end{tabular}
\label{tab:03}
\end{table}

\begin{table}

\caption{
  Sensitivity of stellar parameters to changes of equivalent widths for two stars at the
  cool and hot ends of our sample.}
\begin{tabular}{lcc}
\hline \hline
                                                                                & HD~111884                & HD~36189                \\        \hline
\multicolumn{3}{c}{Changing EW by a factor 1.03}                                                                        \\        \hline
$\Delta$\Teff\    [K]                             & 76               & -21                \\
$\Delta$\logg\   [dex]                       & 0.2                & 0.0                \\
$\Delta \xi$     [$km s^{-1}$]     & 0.0               & 0.0                \\ 
$\Delta$[Fe/H]                     & 0.09              & 0.06                \\ 
    \hline
\multicolumn{3}{c}{Changing EW by a factor 0.97}                                        \\        \hline
$\Delta$\Teff\   [K]                              & -5               & 14               \\
$\Delta$\logg\  [dex]                        & -0.1                & 0.0                \\
$\Delta \xi$ [$km s^{-1}$]       & 0.0               & -0.1               \\        
$\Delta$[Fe/H]                     & -0.03              & -0.02               \\ 
\hline
\hline
\end{tabular}
\label{tab:04}

\end{table}

\section{ Stellar parameters}

\begin{figure}
\centering
\resizebox{\hsize}{!}{\includegraphics{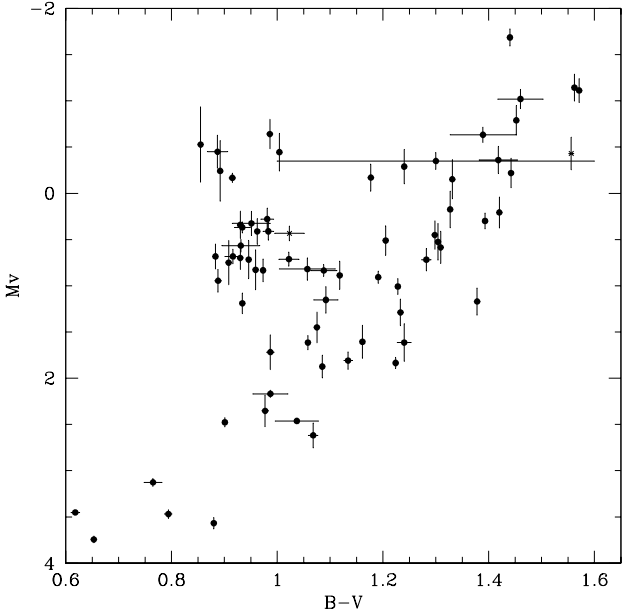}}
\caption{
  Our sample in the color-absolute magnitude diagram.
  Error bars in $M_V$ are derived from the parallax error of the Hipparcos catalog
  ($\sigma_{M_V}=2.17\sigma_{\pi}/\pi$), whereas errors in \bv\ are assumed to be
  the Hipparcos value or $\sigma_{(B\!-\!V)}=0.005$~mag, whichever is larger.
  The few stars with large \bv\ error bars are either known variables or
  -- as for HD~124822, with 0.3~mag-wide error bars --
  suspicious entries in the Hipparcos catalog.
}
\label{cmd}
\end{figure}

The bulk of our sample consists of giants and subgiants in the Hipparcos and Tycho
catalogues (ESA 1997) with parallaxes given with an accuracy better than 10\% .
This means that absolute $M_V$ magnitudes are known with an accuracy of
$\sigma_{M_V}<0.21$~mag, whereas the apparent \bv\ colours (in the Johnson system)
are also known quite precisely.
The typical \bv\ error value given in the catalog is $\la0.005$~mag
except for a few stars classified as variables.
Moreover, reddening is expected to be negligible for this nearby sample,
so that we could initially assume $\bv_0=\bv$.
Fig.~\ref{cmd} shows the sample in the $M_V$ vs. $\bv$ diagram, with error bars included.

Giant stars are well known to suffer the so-called age--metallicity degeneracy:
old metal-poor stars occupy the same region of the color--absolute magnitude diagram
(CMD) as young metal-rich objects.
By having measured the metallicity of our sample stars, it should be possible to
resolve this degeneracy and to estimate stellar ages from the position in the CMD.
Some degeneracy will still remain, for instance we cannot easily distinguish between
first-ascent RGB and post He-flash stars, or between RGB and early-AGB stars.

\subsection{Method for parameter estimation}

We implemented a method to derive 
the most likely intrinsic properties of a star by means of a comparison with a library of
theoretical stellar isochrones, based on the ones by Girardi et al. (2000)
and using the transformations to $BV$ photometry described in Girardi et al. (2002).
We adopt a slightly modified version of the Bayesian estimation method idealized by
J\o rgensen \& Lindegren (2005, see also Nordstr\"om et al. 2004),
which is designed to avoid statistical biases and to take error estimates of all observed
quantities into consideration.
Our goal is the derivation of complete probability distribution functions (PDF)
separately for each stellar property $x$ under study -- where $x$ can be for instance
the stellar age $t$, mass $M$, surface gravity $g$, radius $R$, etc.
The method works as follows:
 
Given a star observed to have $M_V\pm\sigma_{M_V}$, $\Teff\pm\sigma_{T_{\rm eff}}$, and
$[Fe/H]\pm\sigma_{\rm [Fe/H]}$,
\benu
        \item         we consider a small section of an isochrone of metallicity $\feh'$ and
                        age $t'$, corresponding to an interval of initial masses $[\mini^1,\mini^2]$
                        and with mean properties $M_V'$, $\Teff'$ and $x'$;  
        \item         we compute the probability of the observed star to belong to this section
\beqa
        P^{12}(x') & \propto & \int_{\mini^1}^{\mini^2} \phi(\mini) \diff\mini 
                                \exp \left[ -\frac{(\mv-\mv')^2}{\sigma_{\mv}^2} \right.
                                \label{prob12} \\ & &
                                \left. -\frac{(\log\Teff-\log\Teff')^2}{\sigma_{\log T_{\rm eff}}^2}
                \right] , \nonumber
\eeqa
                        where the first term represents the relative number of stars populating
                        the $[\mini^1,\mini^2]$ interval according to the initial mass function
                        $\phi(\mini)$, and the second term represents the probability
                        that the observed \mv\ and \Teff\ correspond to the theoretical values,
                        for the case that the observational errors have Gaussian distributions;
        \item        we sum over $P^{12}(x')$ to obtain a cumulative histogram of $P(x)$;
        \item        we integrate over the entire isochrone;
        \item         we loop over all possible $\feh'$ values, now weighting the $P^{12}(x')$ terms
                        with a Gaussian of mean \feh\ and standard deviation $\sigma_{\rm [Fe/H]}$;
        \item        we loop over all possible $t'$ values, weighting the $P^{12}(x')$ terms with
                        a flat distribution of ages;
        \item         we plot the cumulative PDFs $P(x)$ and compute their basic
                        statistical parameters like mean, median, variance, etc.
\eenu
The oldest adopted stellar age is 12~Gyr, corresponding to an initial mass of about 0.9~\Msun.
Stellar masses as low as 0.7~\Msun\ may be present in our PDFs, since mass loss along the
RGB is taken into account in the Girardi et al. (2000) isochrones.

The method implicitly assumes that theoretical models provide a reliable description of the
way stars of different mass, metallicity, and evolutionary stage distribute along the
red giant region of the CMD.
This assumption is reasonable considering the wide use -- and consequent testing --
of these models in the interpretation of star cluster data.
Also, this assumption can be partially verified {\em a posteriori}, by means of a few
checks  discussed below.

We take the logarithms of the age $t$, mass $M$, surface radius $R$ and gravity $g$,
together with the colour $\bvo$, as the parameters $x$ to be determined by our analysis.
The reason to deal with logarithms is that their changes scale more or less linearly
with changes in our basic observables -- namely the absolute magnitude,
$\log\Teff$, and \feh.
This choice of variables is expected to lead, at least in the simplest cases, to almost
symmetric and Gaussian-like PDFs.

In comparison with Nordstr\"om et al.'s (2004) work, the mathematical formulation we adopt
is essentially the same, except that
\bite
        \item        we use $\log\Teff$ instead of \Teff\ in Eq.~\protect\ref{prob12},
                for the reason stated above,
        \item         we do not apply any correction for the $\alpha$-enhancement of metal-poor
                stars, hence considering all stars in our sample to have scaled solar metallicities.
                In fact, few stars in our sample are expected to exhibit $\alpha$-enhancement
                at any significant level.
        \item         we extend the method to the derivation of several stellar parameters,
                whereas Nordstr\"om et al. (2004) were limited to age and mass,
        \item         we apply the method to red giants with spectroscopic metallicities,
                whereas Nordstr\"om et al. (2004) have a sample composed mostly of main sequence
                stars with metallicities derived from $uvby\beta$ photometry.
\eite
Therefore, the empirical tests and insights provided by our sample will 
be different from, and hopefully complementary to, those already presented by other authors.

We  applied this method earlier, but assuming $\bvo$ as the observable and \Teff\ as a
parameter to be searched for.
This means that \Teff\ is replaced by \bv\ in Eq.~\protect\ref{prob12}, and vice-versa.
Results of this early experiment are presented in Girardi et al. (2006),
and are equivalent to the ones presented here with respect to the estimates of masses,
ages, \logg\ and radii.
The use of a colour instead of the observed \Teff\ in Eq.~\protect\ref{prob12} could
be an interesting alternative for studies of red giants for which spectroscopic \Teff\
is not accurate enough -- e.g. because the available spectra cover a too limited range
and the \feh\ analysis is based on too small a number spectral lines.
\Teff\ could then be derived from the photometry via a \Teff--color relation.
We refer to Girardi et al. (2006) for a discussion of this point.

The implementation of this method will soon be made available, via an
interactive web form, at the URL {\tt http://web.oapd.inaf.it/lgirardi} 

\subsection{Examples of PDFs}
\label{exa_pdf}

\begin{figure*}
\begin{minipage}{0.48\textwidth}
\resizebox{\hsize}{!}{\includegraphics{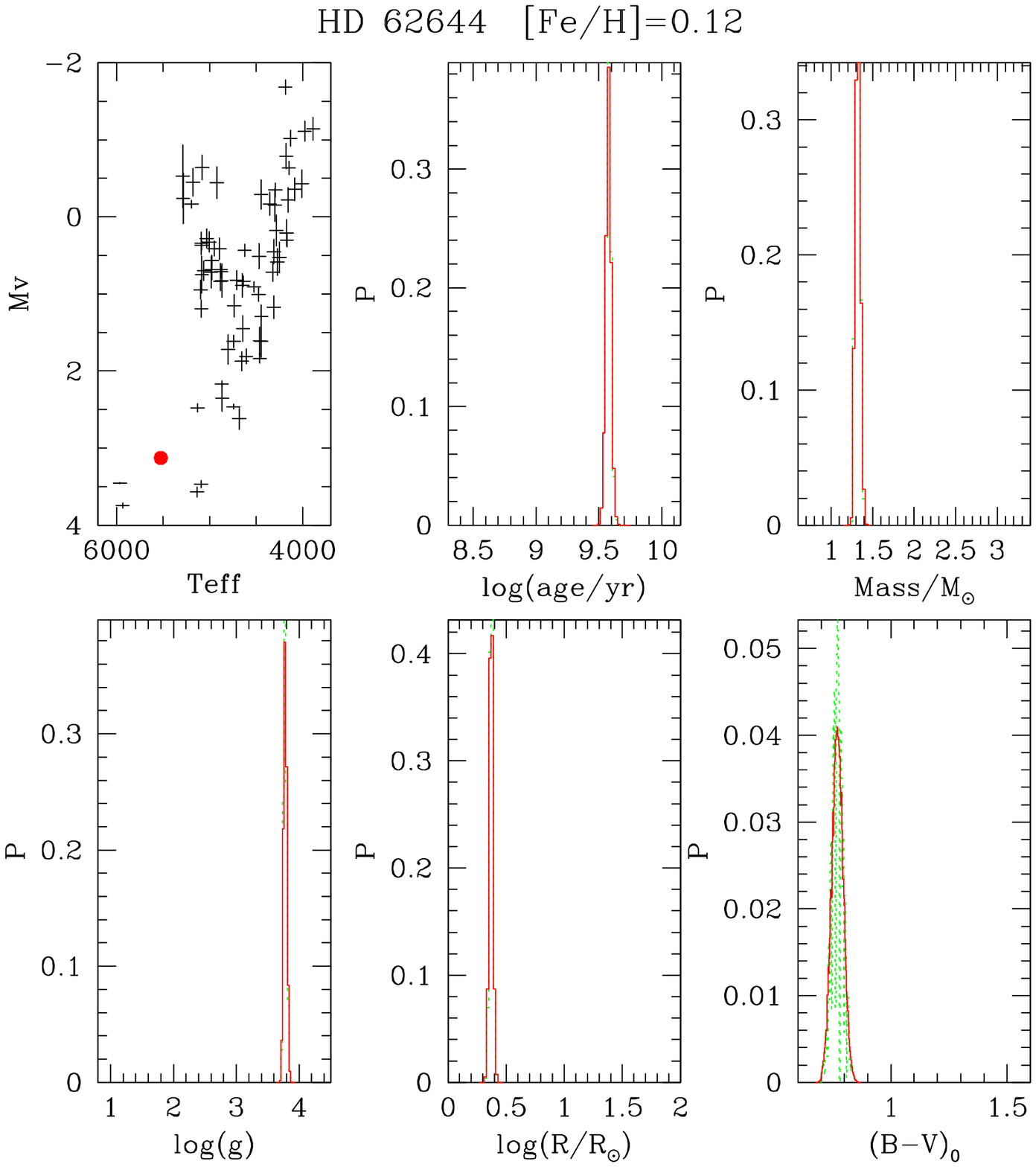}}
\resizebox{\hsize}{!}{\includegraphics{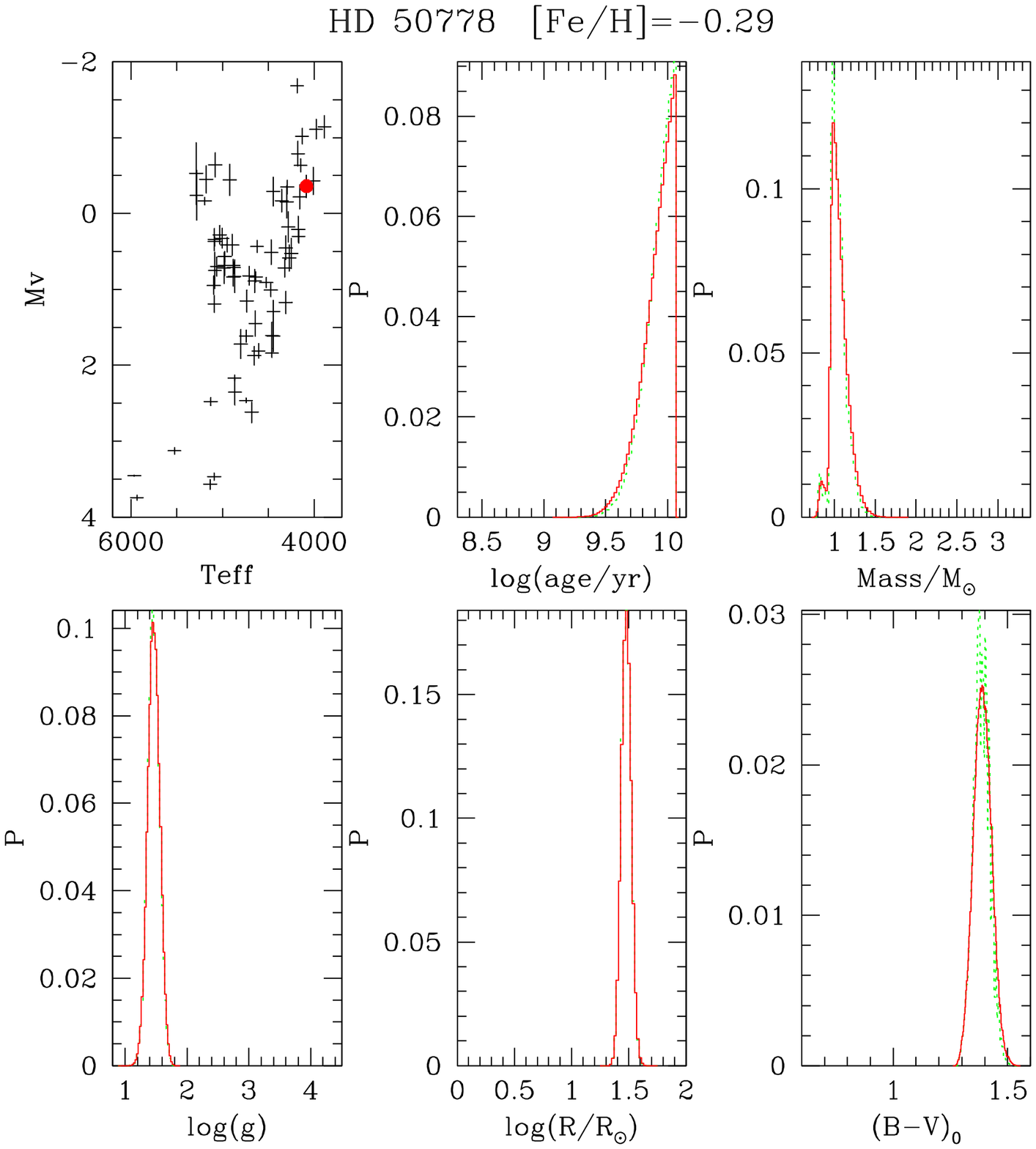}}
\end{minipage}
\begin{minipage}{0.48\textwidth}
\resizebox{\hsize}{!}{\includegraphics{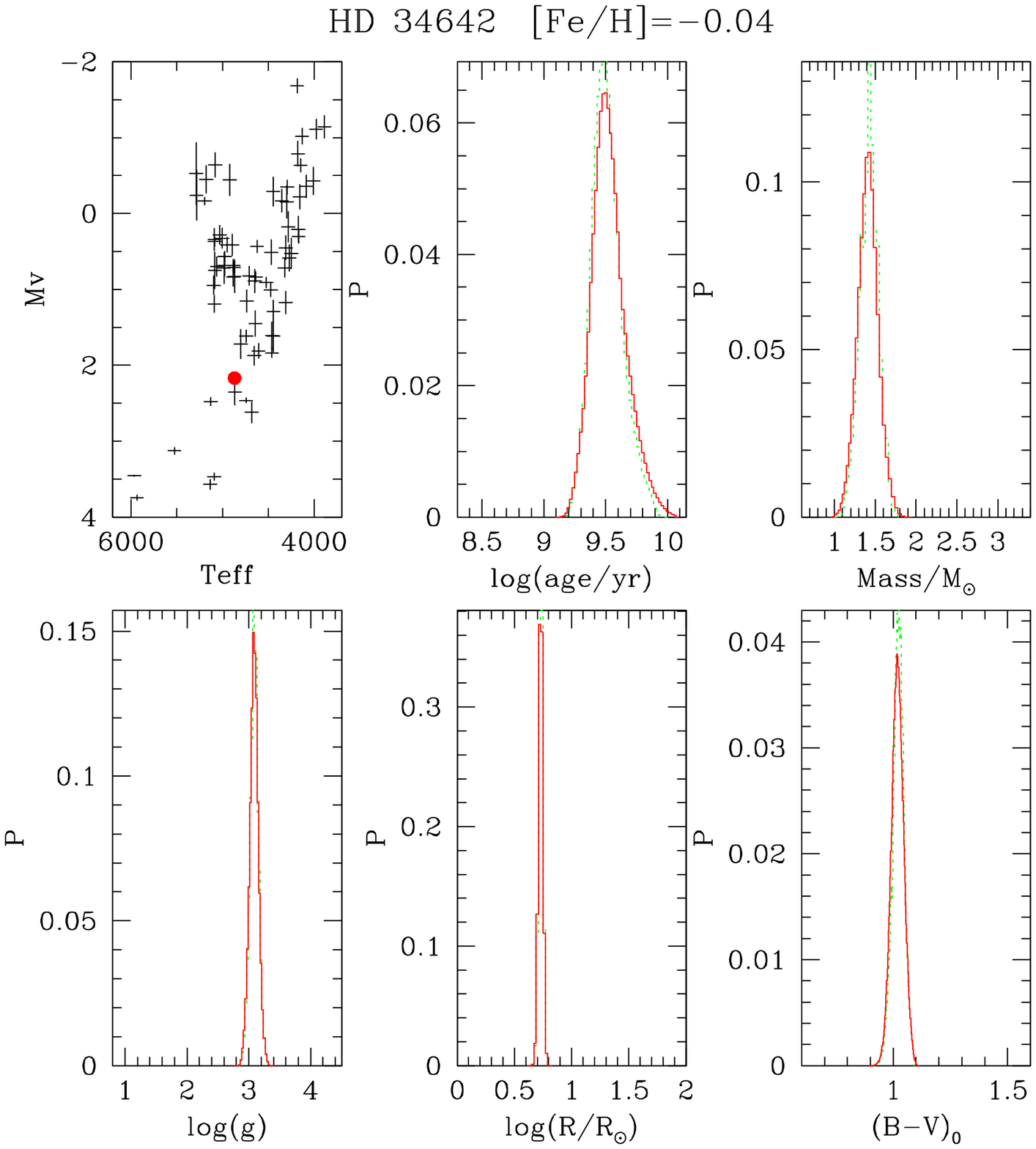}}
\resizebox{\hsize}{!}{\includegraphics{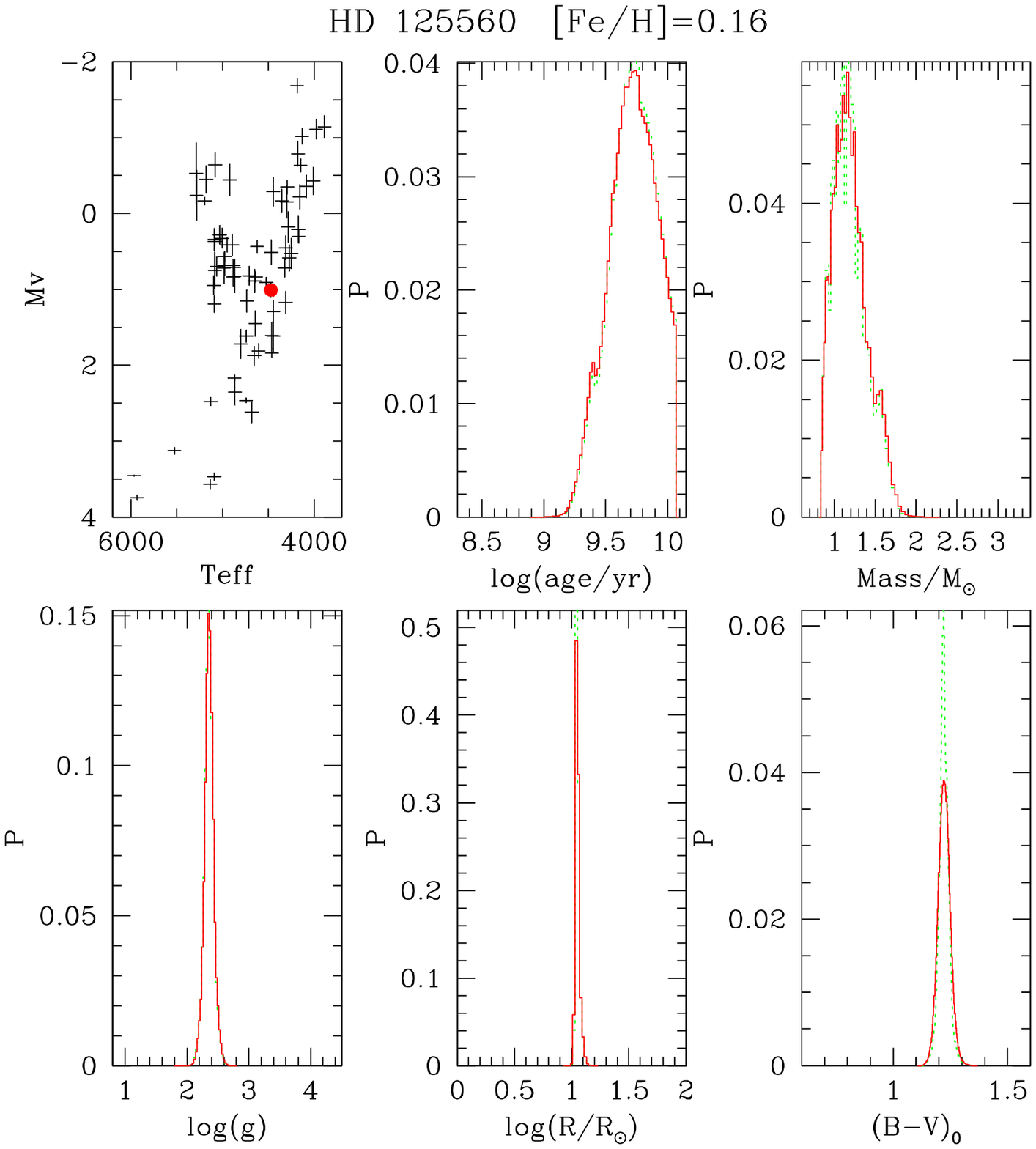}}
\end{minipage}
\caption{
        Examples of probability distribution functions (PDFs), illustrating the majority
        of well-behaved cases in our sample for which good mass and age estimates are possible.
        For each star, one panel presents the position in the HR diagram (dot).
        The five remaining panels show the PDF for $\log t$, $M$, \logg, $\log R$, and \bv.
        The dotted lines show the PDF assuming no error in the \feh\ determination;
        in this case, the PDF width is mostly determined by the parallax errors.
        The solid lines show the slightly broader PDF obtained by assuming an internal
        $\sigma_{\rm [Fe/H]}$ of 0.05~dex.
        HD~34642 and HD~62644 (upper panels) represent stars in the lower part of the RGB
        for which single and well defined peaks in the PDF are typical.
        HD~50778 and HD~125560 (bottom panels), being located in the upper part of the RGB,
        exhibit small secondary peaks in the mass and age PDFs, which represent a small
        probability that the star belongs to an evolutionary phase different from the one
        causing the main peak.
}
\label{pdf1}
\end{figure*}

\begin{figure*}
\begin{minipage}{0.48\textwidth}
\resizebox{\hsize}{!}{\includegraphics{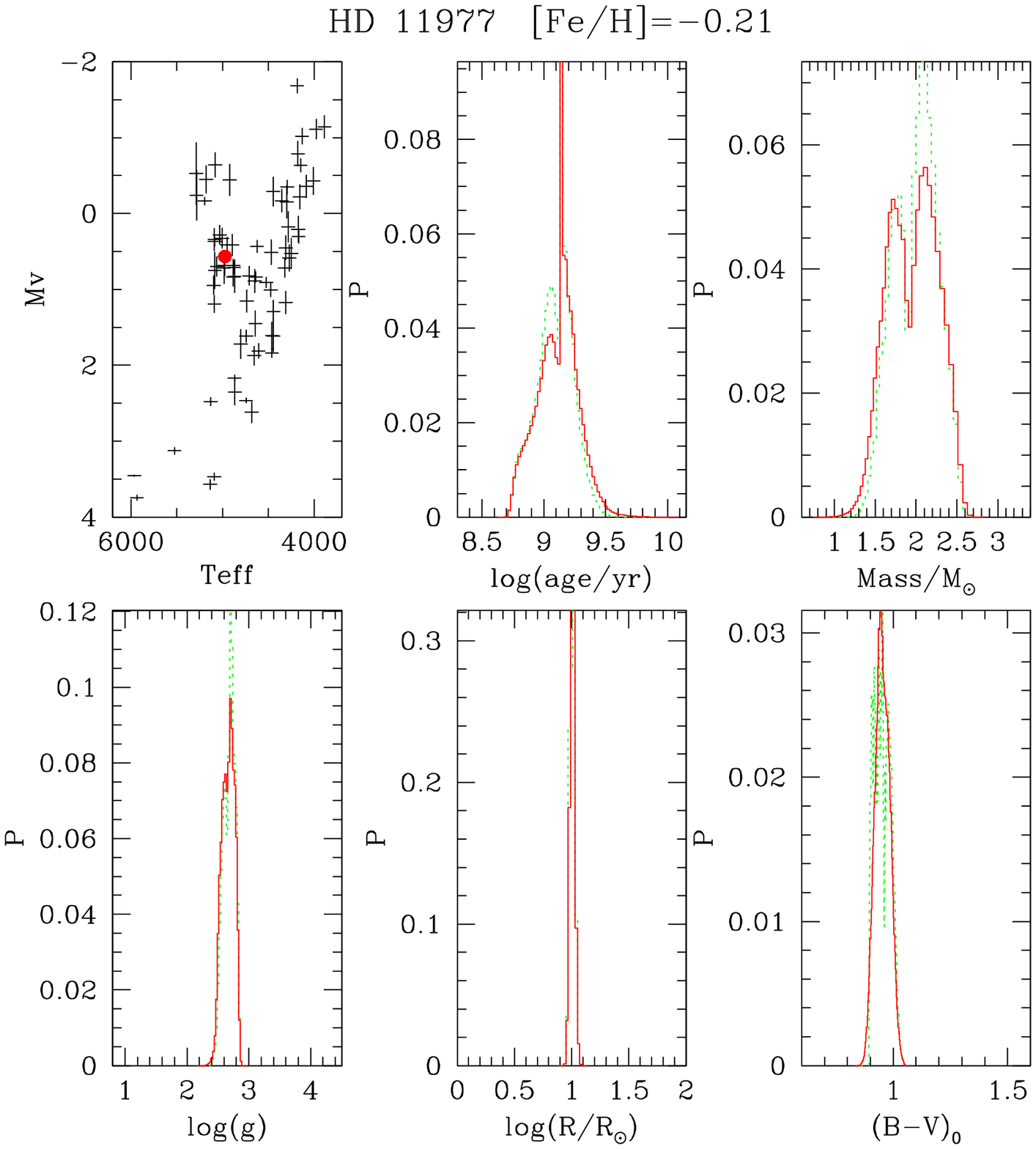}}
\end{minipage}
\begin{minipage}{0.48\textwidth}
\resizebox{\hsize}{!}{\includegraphics{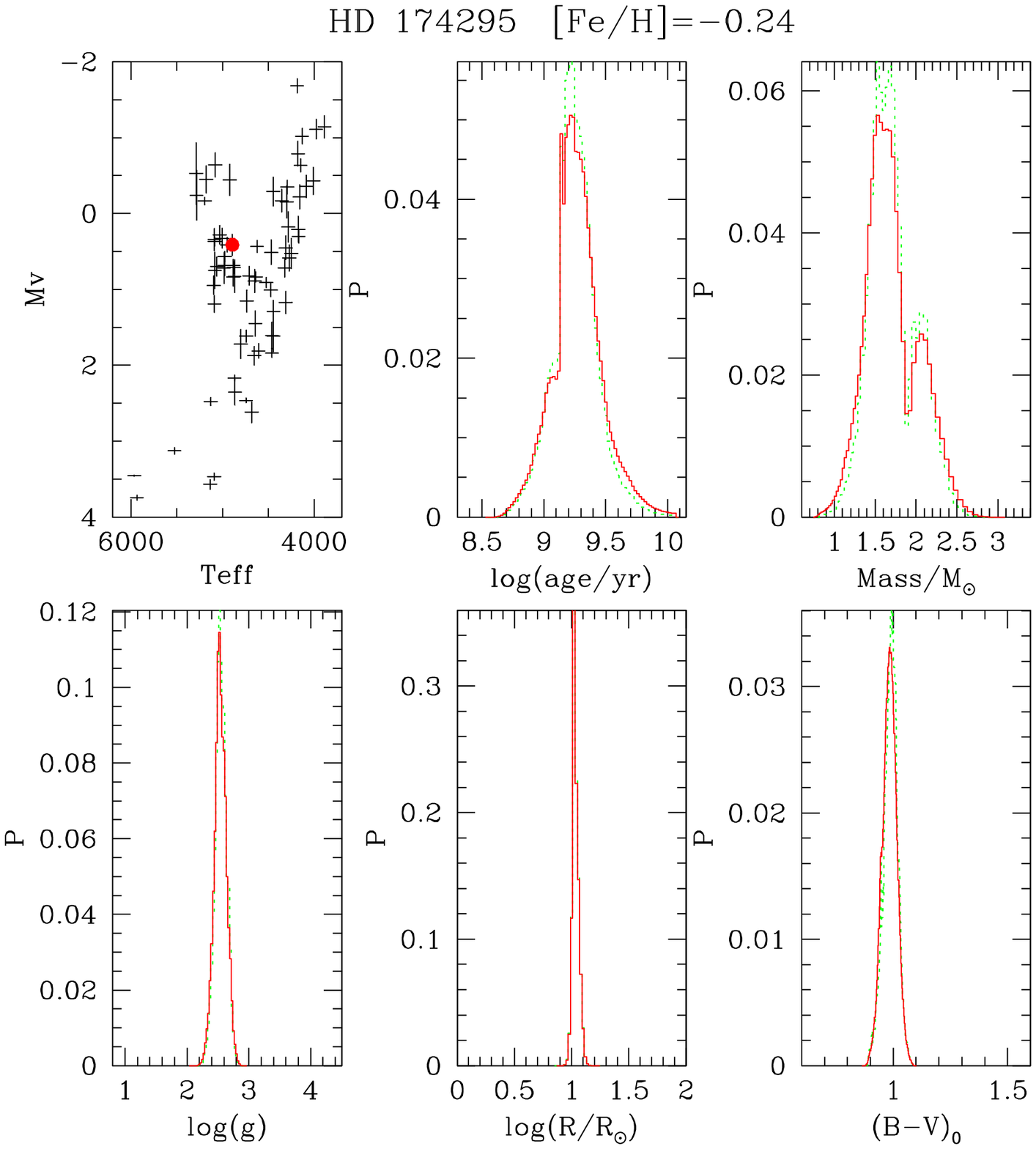}}
\end{minipage}
\caption{
        Same as Fig.~\protect\ref{pdf1}, illustrating two cases for which the estimate
        of stellar parameters is not unique.
        Both stars are in the red clump region of the HR diagram.
        Their parameters can be well reproduced by clump stars of different masses/ages,
        which are represented by the two main peaks in their PDFs.
}
\label{pdf2}
\end{figure*}

Since one of the main achievements of this work is the application of the
J\o rgensen \& Lindegren (2005) method to a sample of giants, it is important
to illustrate the different situations we encounter for stars along the RGB.
Figures~\ref{pdf1} and \ref{pdf2} provide a few examples of PDFs
derived for stars in different positions in the CMD.
They illustrate the typical cases of PDFs with single, well-defined peaks
(Fig.~\ref{pdf1}) as well as some cases for which the parameters cannot
uniquely be determined (Fig.~\ref{pdf2}).

Two cases for which the parameter determination works excellently are
HD~62644 and HD~34642 (Fig.~\ref{pdf1}, upper half).
The PDFs present a single prominent peak and a modest dispersion for all
estimated parameters, permitting a clear identification of the most likely
parameter value by the mean and its uncertainty as defined by the standard error.
The errors in determining mass and age are however much
smaller for HD~62644 than for HD~34642, even if these two
stars have a similar (and very small) parallax error.
This is so because HD~62644 is still a subgiant, whereas HD~34642 is an
authentic red giant.
The evolutionary tracks for different masses for HD~62644 run horizontally and well
separated in the CMD, whereas for HD~34642 they are more vertical and much closer.
The same input error bars will in general produce larger output errors for
a giant than for a subgiant.

One notices that the errors in parameter determination for these two cases
originate primarily from the error in the Hipparcos parallax, and are just
slightly increased by the effect of a 0.05~dex error in \feh.
In fact, the \feh\ error involves the age-metallicity degeneracy, especially for the red giant HD~34642.
It is also worth noticing that parameters like $g$, \Teff, and $R$
depend slightly on the measured metallicity and its error.
Acceptable values for these parameters could have been derived
assuming a fixed -- say equal to solar -- metallicity.
However, our method has the advantage of accounting fully for the subtle
variations of the colour and bolometric correction scales with metallicity,
thereby slightly reducing the final errors in the parameter determination
when the observed \feh\ is taken into account.
In contrast, for estimating $t$ and $M$, a proper evaluation of the metallicity
and its error turns out to be absolutely necessary.

Other well-behaved cases are HD~50778 and HD~125560 (Fig.~\ref{pdf1}),
which are located in the upper part of the RGB, one well above and the other
close to the red clump region.
A robust mass and age determination is possible for these stars, but their mass
PDFs show a faint secondary peak close to a prominent primary peak.
For HD~50778, the primary peak corresponds to $\sim0.95$~\Msun\ for a star in the
phase of first-ascent RGB, whereas the secondary peak corresponds to the early-AGB
phase of a $0.7$~\Msun\ star.
There is no way to distinguish {\em a priori} between these two
evolutionary phases from our observations, but fortunately the first-ascent RGB case
turns out to be much more likely (due to its longer evolutionary timescale)
than the early-AGB phase.
In the case of HD~125560, the primary peak in the PDF corresponds to a red clump
(core He-burning) phase of a 1.1~\Msun\ star, whereas the secondary peak corresponds
to a 1.6~\Msun\ star in the first-ascent RGB phase.
Similar results, with the presence of small secondary peaks in the PDF, are common
for stars in the upper part of the CMD.

Parameter estimation is much more difficult for stars like HD~11977 and HD~174295
(Fig.~\ref{pdf2}).
These stars are in the middle of the most degenerate region of the CMD,
namely in the ``loop'' region of red clump stars of different masses.
As illustrated in Girardi et al. (1998, their figure 1),
core He-burning stars with the same metallicity and different mass form a compact loop
which starts for low masses in the blue end as an extension of the horizontal branch,
reaches its reddest colour as the mass increases and then turns back into the blue direction.
The luminosity along the same mass sequence first increases slowly by some tenths of a
magnitude, decreases sharply by 0.5~mag at about 2~\Msun, and turns towards much higher
luminosities as the mass increases further.
Such a complex pattern in the CMD implies that for some stars in the middle of such
loops, two mass (and age) values may become similarly likely, resulting in bi-modal PDFs.
Moreover, the shapes of such PDFs may become sensitive to even small changes in \feh,
thus increasing further the uncertainty of the determination of the stellar parameters.
Ill-behaved cases like the ones illustrated in Fig.~\ref{pdf2} account for less than
a fifth of our sample.

\subsection{Results and checks}

Results for all our sample stars are presented in columns 6 to 17 of Table \ref{tabelao}.
Except for \bvo, we have determined mean values and standard errors using PDFs of
logarithmic quantities, and subsequently converted these values into linear scales.
We have done so because linear quantities (age, mass, radii, etc.) are more commonly
used and are considered to be more intuitive than the corresponding logarithms.
Whenever possible, however, we will use the original error bars obtained with
the logarithmic scale.

We have applied a few checks to what extent
our method for parameter estimation is reliable.

\subsubsection{``Colour excess'' $\bv-\bvo$ }

\begin{figure}
\resizebox{\hsize}{!}{\includegraphics{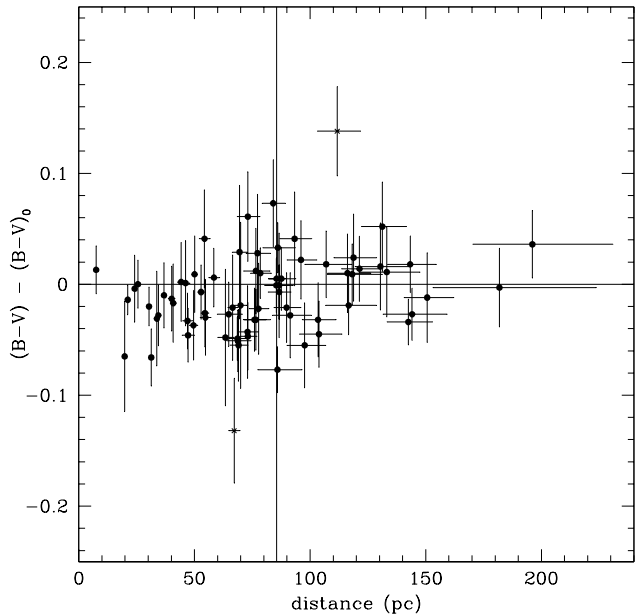}}
\caption{
    Difference between the observed \bv\ values, and the \bvo\ values derived from
        the PDF method, as a function of distance from the Sun.
        Notice the absence of a clear trend of $\bv-\bvo$ with distance.
        The data points shown as crosses denote the cases which we consider as ``outliers'' and
        which are excluded from our statistical analysis.
}
\label{ebv_dist}
\end{figure}

\begin{figure}
\resizebox{\hsize}{!}{\includegraphics{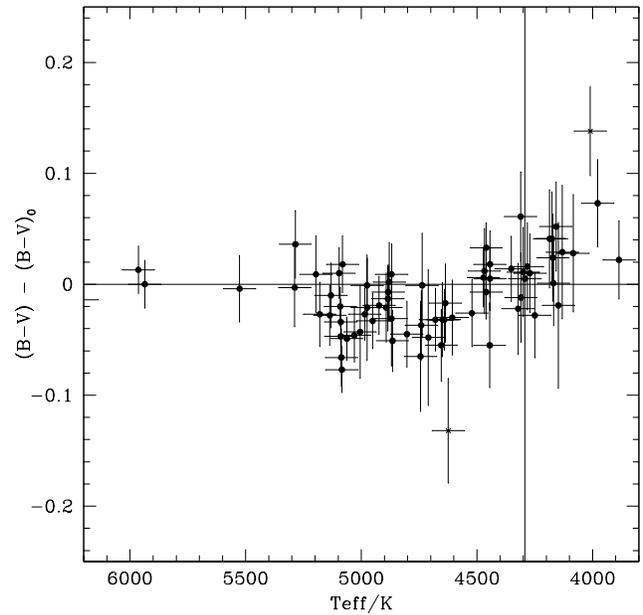}}
\caption{
        The same as Fig.~\protect\ref{ebv_dist}, but as a function of effective temperature.
        Notice that the small differences found for most of the stars (less than 0.05~mag)
        appears to be a function of \Teff.
}
\label{ebv_teff}
\end{figure}

The first test involves a comparison of the estimated intrinsic colours \bvo\ with the
observed colours \bv.
Their differences are presented in Figs.~\ref{ebv_dist} and \ref{ebv_teff} as a function
of distance and \Teff.
Since the \bvo\ values result essentially from the spectroscopic \Teff, error bars in
the individual $\bv-\bvo$ values reflect mostly the 70~K error assumed for \Teff,
except for the few stars for which the \bv\ error of the Hipparcos catalogue
was significant (see also Fig.~\ref{cmd}).

As can be clearly seen in Fig.~\ref{ebv_dist}, there is no marked increase of $\bv-\bvo$
with distance, as can be expected for a sample with distances less than $\sim200$~pc  
and correspondingly little reddening.
Taking the diffuse interstellar absorption of $\diff A_V/\diff r=0.75$ mag/kpc
(Lyng\aa\ 1982), one expects a color excess of just $\ebv=0.05$~mag at 200~pc.
This expectation is consistent with our $\bv-\bvo$ data, although comparable
with their dispersion.

Another aspect to notice in Fig.~\ref{ebv_dist} is the small dispersion of the $\bv-\bvo$
values we obtained for the bulk of the sample stars.
If we disregard two outliers with $|\bv-\bvo|\ga0.1$, we find an unweighted mean
of $\bv-\bvo=-0.009$ with a scatter of $0.031$.
This scatter (excluding outliers) can be considered the typical error of our
PDF method for determining the intrinsic colour of our sample.

Fig.~\ref{ebv_teff} shows how $\bv-\bvo$ depends on \Teff.
There is evidently a correlation between $\bv-\bvo$ and \Teff, with a
minimum difference at $\sim4700$~K and a maximum difference at $\sim4000$~K.
It is likely that this correlation is caused by errors in the theoretical
\Teff--colour relation adopted in the Girardi et al. (2002) isochrones,
which amount to less than 0.05~mag, or equivalently to about 100~K for a given \bvo
\footnote{
        A small contribution to this behaviour with \Teff may be due to modest
        reddening of the most distant stars, which have a tendency towards
        brighter absolute magnitudes and lower \Teff.
}.
In addition, even if our \Teff--colour scale  were perfectly good, \bv\ starts to
become intrinsically a poor \Teff\ indicator for the coolest giants.

Notice that the possible systematic errors of 0.05 mag in our adopted \Teff--colour
relation would imply errors smaller than 0.03 for the $V$-band bolometric corrections
adopted for the same isochrones, which would then be the maximum mismatch between
theoretical and observational \mv\ values.
We conclude that these errors are small enough to be neglected in the present work.

The two outliers with high $|\bv-\bvo|$ in Fig.~\ref{ebv_dist} (HD~22663 and HD~99167) 
can be explained as (1) stars with a significant reddening;
(2) stars for which Hipparcos catalogue has a wrong entry for \bv;
or (3) stars for which our parameter estimation (including \Teff, \feh, and/or \bvo)
substantially failed.
We consider the third alternative as being the most likely one, and hence exclude these
two stars from any of the statistical considerations that follow in this paper.

\subsubsection{Surface gravities}

\begin{figure}
\begin{center}
\resizebox{\hsize}{!}{\includegraphics{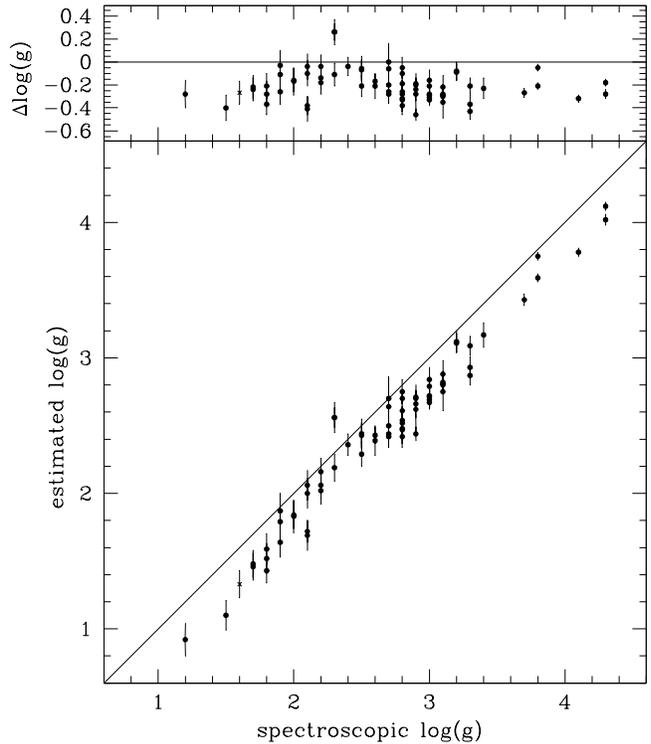}}
\end{center}
\caption{
        Comparison between the \logg\ values derived from the spectra (abcissa) and from
        the photometry by means of our PDF method (ordinate).
        The solid line represents a one-to-one relation. The upper panel shows the differences between estimated and spectrocopic values.
}
\label{logg}
\end{figure}

Another important check is the comparison between our estimated \logg\ values with those
derived independently from spectroscopy (see Sect.~\ref{chemical}), presented in Fig.~\ref{logg}. The PDF-estimated values tend to be systematically
lower than the spectroscopically derived ones.
Again ignoring the two outliers of the previous section, the mean difference
is $-0.20$~dex with a standard deviation of 0.14~dex.
Such an offset would indicate, for instance, that our method underestimates
stellar masses by a factor of about 1.6, which however can be excluded given
our results for the two Hyades giants (see below).
An alternative explanation is that the spectroscopic \logg\ values are simply too high.

The latter interpretation is supported by the consideration that gravity is determined
by imposing ionization balance; this means that the abundance found for the nine Fe~II lines
is the same as the one retrieved for the (more than 70) Fe~I lines.
This procedure implies that spectroscopic gravities depend, in addition to the adopted
line oscillator strengths, to the interplay between the stellar parameters in the
derivation of abundances.
This can be fairly complex in Pop~I giants, where the Fe I vs. Fe II abundance depends
not only on gravity, but also quite strongly on effective temperature.
As an example, see Pasquini et al. (2004, their table 4),
where the dependence of Fe~I and Fe~II on \logg, \Teff\ and $\xi$ is analyzed
for one Pop~I giant and the same set of lines.
A systematic shift of 100~K in \Teff\ would, for instance, produce a 0.2~dex shift
in \logg\ without changing substantially the derived Fe abundance.

The disagreement between the gravity values obtained from spectroscopy and from parallaxes has been known for a long time (e.g., da Silva 1986), and the problem did not 
disappear despite the improvements of models and parallax measurements.
Nilsen et al. (1997) compare the Hipparcos-based gravities with the values obtained from
spectroscopy by several authors and conclude that differences between the two methods
could become larger than a factor two (0.3 in \logg\ ).
This can have various causes, like non-LTE effects on Fe I abundances, or thermal
inhomogeneities.
We conclude that our spectroscopic gravities, like that of other authors, are systematically
overestimated.
Note also that Monaco et al. (2005, their figure 6) find that spectroscopic \logg\ values
correlate with microturbulent velocities $\xi$, in such a way that a systematic error
of just $0.07$~km~s$^{-1}$ in $\xi$ would be sufficient to cause the $-0.20$~dex offset
which we find in \logg.
We point out that the $-0.20$~dex offset in \logg\ would have a negligible effect on our
derived \feh\ values, that are mostly based on the gravity-insensitive Fe~I lines (see section 3).

 
\subsubsection{Stellar radii and apparent stellar diameters}

\begin{figure}
\begin{center}
\resizebox{\hsize}{!}{\includegraphics{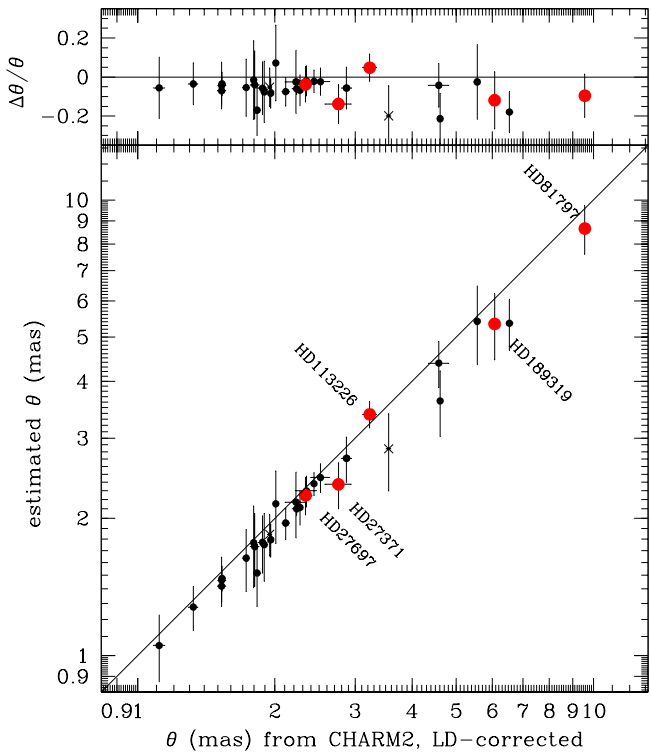}}
\end{center}
\caption{
        Comparison of apparent stellar diameters $\theta$ from the
        CHARM2 catalogue (abcissa) and derived photometrically with
        our PDF method (ordinate).  The entries in CHARM2 were
        corrected by limb darkening as detailed in the text.  The
        solid line represents the one-to-one relation. The large dots
        with HD identification indicate stars with $\theta$
        measurements obtained via lunar occultation and/or LBI
        techniques. The upper panel shows the relative differences
        $\Delta\theta/theta$, i.e., the differences between estimated
        and measured values, divided by the measured $\theta$.
}
\label{theta}
\end{figure}

Another  check regards the stellar radii $R$, which can be
easily converted into apparent stellar diameters $\theta$ using
Hipparcos parallaxes, and then compared to observations.  The observed
apparent diameters $\theta$ are taken from the CHARM2 catalogue
(Richichi et al. 2005), which in most cases consist of indirect
estimates of stellar diameters using fits to spectrophotometric data,
and are available for about one third of our sample.  ``Direct''
diameter determinations are available only for five of our giants,
based on lunar occultation (HD~27371 and HD~27697; the Hyades giants)
and on long baseline interferometry (LBI; HD~27697, HD~81797,
HD~113226, and HD~189319).  In cases for which more than one $\theta$
determination was available, we used either the most accurate one
(i.e. the one with a substantially smaller error) or the most recent
one when tabelled errors were similar.  Finally, we have corrected
all uniform-disk measurements by limb darkening (LD) using the
extensive tables provided by Davis et al. (2000), that are based on Kurucz
(1993ab) model atmospheres. For the 20 sample stars with LD-corrected
diameters in CHARM2, our corrections agree perfectly with those
provided there. The mean LD-correction for these giants is 
 $3.7\pm3.3$~\%, which is well below the $\sim12$~\%\ $1\sigma$
relative error of our individual $\theta$\ estimates.

The comparison between our derived $\theta$ values and the 
LD-corrected CHARM2 diameters is presented in Fig.~\ref{theta}.  Here
again, the comparison is very satisfactory, with the
estimated minus observed difference  $-0.21$~mas
with a scatter of 0.32~mas.  Alternatively, we looked at the fractional
differences $\Delta\theta/\theta_{\rm CHARM2}$ (upper panel of
Fig.~\ref{theta}): its unweighted mean value is of
$\delta\theta/\theta=-0.06$ with a r.m.s. scatter of $0.06$. This
scatter also is well below the fractional error of individual $\theta$
estimates, of $\sigma_\theta/\theta\simeq0.12$ (see the error bars in
the upper panel of Fig.~\ref{theta}), which are largely due to
Hipparcos parallax errors.

A similar level of agreement is obtained for stars with
``direct'' $\theta$ mesurements.  For all the other stars in
Fig.~\ref{theta}, the ``indirect estimates'' based on fits to
spectrophotometric data (and especially infrared data) presented in
CHARM2 correlate very well with our values.  This is a remarkable
result.  It indicates that by using just two visual passbands ($B$ and
$V$ as in the present work) for giants of known distance, \Teff\ and
metallicity, it is possible to obtain diameter estimates of a quality
similar to that obtained by more sophisticated methods based
on multi-band spectrophotometry.

Our results indicate a very successful and robust
estimation of the stellar parameters \bvo, \logg, and $R$, for the
bulk of our sample stars.

\subsubsection{Ages and masses}

\begin{figure}
\begin{center}
\resizebox{\hsize}{!}{\includegraphics{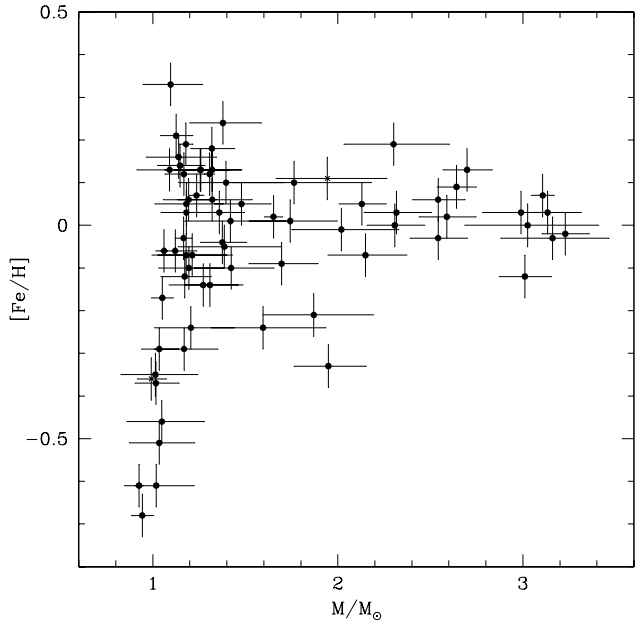}}
\end{center}
\caption{Mass--metallicity relation for our sample stars.}
\label{mmr}
\end{figure}

\begin{figure}
\begin{center}
\resizebox{\hsize}{!}{\includegraphics{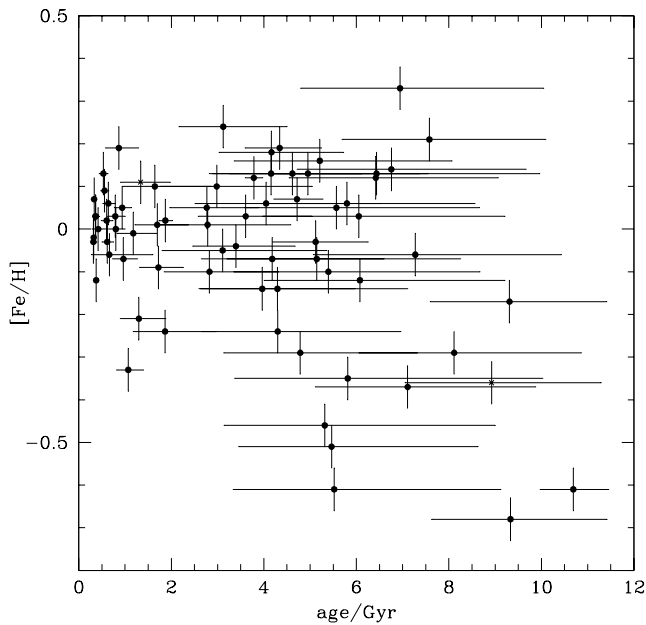}}
\end{center}
\caption{Age--metallicity relation for our sample stars.}
\label{amr}
\end{figure}

With respect to the parameters $t$ and $M$, the few checks at our disposal
address the correlations with \feh, and the results for the two Hyades giants in our sample.
Fig.~\ref{mmr} presents the mass--metalicity plot.
It shows a clear pattern: low-metallicty giants (with $\feh<-0.4$) are present
only among the stars with the lowest masses.
All stars with $M>1.2$~\Msun\ are characterised by a mean solar metallicity (0.00~dex)
and a small standard deviation of 0.12~dex.
The same data, when presented as an age--metallicity plot in Fig.~\ref{amr},
indicates a more scattered pattern but with the similar indication of metal-poor stars
being present with an age above $\sim5$~Gyr.
The present data points to an age--metallicity relation with a large scatter for the
highest ages.
This scatter is consistent with other results in the literature, at least for
the highest ages (see Sect.~\ref{more_amr}).

Our sample contains two Hyades giants for which good-quality
age and mass estimates are available:
HD~27371 with $t=0.53\pm0.09$~Gyr, $M=2.70\pm0.13$~\Msun, and
HD~27697 with $t=0.67\pm0.13$~Gyr, $M=2.54\pm0.14$~\Msun.
The Hyades turn-off age, as derived from models with overshooting,
is $0.625\pm0.05$~Gyr (Perryman et al. 1998).\footnote{
        Ages close to 0.63~Gyr are also indicated by the Hyades binaries
        V~818~Tauri, 51~Tauri, and $\theta^2$~Tauri (Lastennet et al. 1999,
        in their table 6), whereas the white dwarfs provide a lower age limit
        of $0.3\pm0.03$~Gyr (Weidemann et al. 1992).}
This value is consistent to within $1\sigma$ with our estimated ages
for HD~27697 and HD~27371.

Although the error in our age estimates for individual stars is unconfortably
large compared to the typical error of cluster turn-off ages,
the pair of Hyades giants provides the best evidence that the
J\o rgensen \& Lindegren (2005) method works well for estimating the ages of giants.
Unfortunately, additional checks using other clusters are apparently impossible
at this moment.
Although other well-studied clusters with excellent turn-off ages exist
(e.g. M67 and Praesepe), they do not belong to our sample and have significantly
smaller parallaxes.

To summarize, we conclude that our method has provided excellent determinations
of stellar parameters, especially for \bvo, \logg, $R$ and $\theta$ for which
the uncertainties of the method were intrinsically low (with a few exceptions),
and has been confirmed by independent data.
For the stellar ages and masses, however, our determinations turn out to be
intrinsically more uncertain, as demonstrated by the larger error bars we obtained.
Although we have some indication from the Hyades giants that our age scale
is not very inaccurate, another independent check with other mass and age
data would be desireable.

\section{More about the age--metallicity relation}
\label{more_amr}

The Solar Neighbourhood age-metallicity relation (AMR) provides basic information
about the chemical evolution of the Milky Way's disk with time, and has for
long been used to constrain evolutionary models of our Galaxy.
We refer to Carraro et al. (1998), Feltzing et al. (2001) and Nordstr\"om et al.
(2004) for recent determinations of the AMR and a general discussion of its properties.

Since we have derived the AMR from a completely new sample and use
a relatively new method, it is important to illustrate how our results
compare with other determinations.
Note also that our sample is appropriate to do this because the stars were
not chosen using any criterion of age, abundance or galatic velocity.  

\begin{figure}
\begin{center}
\resizebox{\hsize}{!}{\includegraphics{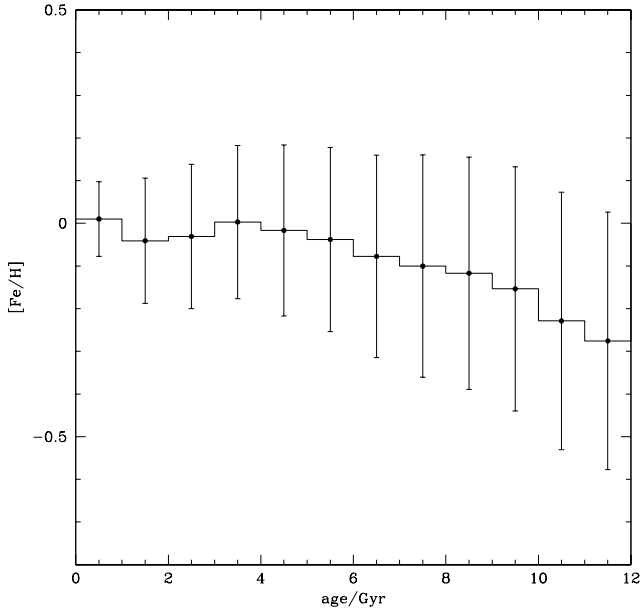}}
\end{center}
\caption{
        Age--metallicity relation for 1~Gyr age bins (see Table~\protect\ref{tab_amr}).
}
\label{fig_amr}
\end{figure}

First, we transform our AMR data of Fig.~\ref{amr} to a more simple
function of age.
To do so, for each age bin $\Delta t$, we determine its cumulative \feh\ by
adding the measured \feh\ of each star weighted by its probability of
belonging to $\Delta t$.
This probability is given by the age PDF of each star, integrated over
the $\Delta t$ interval.
We obtain a cumulative \feh\ distribution for each age bin, from which we
derive the mean \feh\ value and dispersion.
Since the metallicity value of a single star is spread over several
age bins (just as for the age PDF), the effective number of
stars per bin, $\langle N\rangle$, may be a fractional number.
We choose age bins wide enough to provide $\langle N\rangle>1$ for all ages.
Note that the data points obtained this way for different age bins are not independent.
The results are presented in Fig.~\ref{fig_amr} and Table~\ref{tab_amr},
where the mean metallicity and its dispersion is shown as a function of age.

\begin{table}
\caption{
        \label{tab_amr}
        Age--metallicity relation derived from our sample for 1-Gyr wide age bins.
}
\begin{tabular}{r|rrr}
\hline\hline
$\langle t\rangle$ (Gyr) & $\langle \feh\rangle$ &
$\sigma_{\rm [Fe/H]}$ & $\langle N\rangle$ \\
\hline
0.5 & $ 0.01$ & 0.09 & 17.1 \\
1.5 & $ -0.04$ & 0.15 & 10.5 \\
2.5 & $ -0.03$ & 0.17 & 6.1 \\
3.5 & $ 0.00$ & 0.18 & 9.1 \\
4.5 & $ -0.02$ & 0.20 & 5.3 \\
5.5 & $ -0.04$ & 0.22 & 5.3 \\
6.5 & $ -0.08$ & 0.24 & 4.7 \\
7.5 & $ -0.10$ & 0.26 & 3.4 \\
8.5 & $ -0.12$ & 0.27 & 2.3 \\
9.5 & $ -0.15$ & 0.29 & 2.5 \\
10.5 & $ -0.23$ & 0.30 & 4.2 \\
11.5 & $ -0.28$ & 0.30 & 1.6 \\
\hline
\end{tabular}
\end{table}

Our results present a few notable characteristics:
\benu
\item         The AMR  relatively flat up to the largest ages.
                From its present-day solar's value, the mean metallicity has dropped
                to just $-0.2$~dex at an age of 12~Gyr.
                This result appears to be in qualitative agreement with Nordstr\"om et al. (2004).
                Other authors arrive at a somewhat lower mean metallicity at the oldest ages
                (e.g., Carraro et al., 1998; Rocha-Pinto et al., 2000; Reddy et al., 2003).
                As discussed by Nordstr\"om et al. (2004) and Pont \& Eyer (2004),
                selection against old metal-rich dwarfs may have contributed to the steeper
                decline of \feh\ with age found in many of the previous studies.
                It is likely that such selection effects are absent in our sample of giants.

\item        The metallicity dispersion tends to increase with age, so that one finds a
                large \feh\ dispersion of about 0.3~dex ($1\sigma$) for the highest ages.
                Part of this trend may result from the increase in the age error with age
                (evident in Fig.~\ref{amr}), which causes stars of very different ages to
                contribute to the mean \feh\ of the same age bins.
                It seems clear from our data that for ages larger than about 4 Gyr,
                the \feh\ dispersion becomes considerably larger than the observational errors.
                This is in fairly good agreement with the results of other authors.

\item         For all ages lower than about 4 Gyr, we find that the
                \feh\ dispersion is comparable to the typical observational error.
                The most striking result hoewever is  given by the 0 to 1-Gyr age bin,
                which is very well populated (17.1 stars) and where age errors are
                typically very low so that the confusion with other age bins is
                practically absent.
                In this case, the \feh\ dispersion is 0.09 dex, and compares well with
                the observational \feh\ errors of 0.05~dex (internal) and 0.1~dex (external).
                This result is in contradiction into the claims by most authors
                (Carraro et al., 1998; Feltzing et al., 2001; Nordstr\"om et al., 2004, etc), who find large \feh\ spread at {\em all} ages.
\eenu

Why do our results differ from other authors, in particular for the youngest stars?
The main difference is likely to result from the different kinds of stars investigated
in the mentioned above studies.
We study only giants whereas most authors use field dwarfs and a few subgiants.
As we have demonstrated, the age determination for the youngest giants appears to be
quite reliable.
The same may not be true for dwarfs which, being located close to the main sequence,
are in a more degenerate region of the HR diagram.
The results of Edvardsson et al (1993) and Nordstr\"om et al. (2004),
which represent the best age estimates for a limited section of the main sequence,
report considerable uncertainities for the ages of most of their stars.
In addition, their \feh\ determination is based on $ubvy\beta$ photometry, which
further limits their sample to an even smaller section of the main sequence.
These problems do not exist for giants with spectroscopic \feh\ and \Teff\ determinations.
Reliable measurements can be obtained all along the RGB, and the worst problems
appear not to be related to selection effects, but to the intrinsic age errors illustrated in
Sect.~\ref{exa_pdf}.
We conclude that giants may be better targets for the study of the Solar Neighbourhood
AMR than dwarfs.
As an alternative, subgiants may be even better because they always provide reliable
age estimates, as illustrated in Fig.~\ref{pdf1}, and as already explored by Thor\'en et al. (2004).

The interpretation of our AMR result  indicates that stars in the
Solar neighbourhood are formed from interstellar matter of quite homogeneous chemical composition.
As we observe older stars, we start sampling stars born in different Galactic locations\footnote{
        Different Galactic locations mean different radial positions in the thin disk.
        The oldest age bins may also contain a few thick disk and halo stars.
}, and hence we see a more complex mixture of chemical composition.

\section{Discussion}

\subsection{RV variation }

\begin{figure}
\begin{center}
\resizebox{\hsize}{!}{\includegraphics{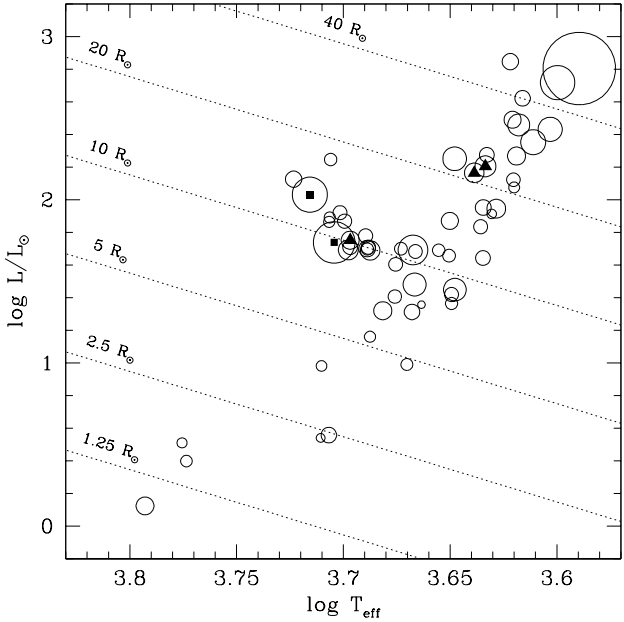}}
\end{center}
\caption{
        HR diagram of our star sample, excluding 11 binaries.
        Stellar parameters were taken from Table~\protect\ref{tabelao}.
        Each circle diameter is proportional to $\sigma_{\rm RV}$,
        ranging from $\sim30$ to 280~m\,s$^{-1}$.
        Stars that host substellar companions are marked with full squares
        for suspected brown dwarfs and full triangles for suspected giant planets.
}
\label{hrd}
\end{figure}

Setiawan et al. (2004) have studied the trend of radial velocity (RV) variability
-- determined by the standard deviation $\sigma_{\rm RV}$ -- with $M_V$,
detecting an increase of variability along the RGB.
We can now do the same for other stellar parameters.
Figure~\ref{hrd} shows our star sample in the HR diagram obtained from the information
of Table~\ref{tabelao}. We note that the $\sigma_{RV}$ represents the total standard deviation
without regard to the timescales involved (i.e. short- versus long
term variations).
The 11 binaries of the sample which are reported by Setiawan et al. (2004) are excluded
from this plot.
There is a trend of increasing $\sigma_{\rm RV}$ with position along the RGB,
which becomes more evident if we exclude the stars that host sub-stellar companion
candidates (marked with different symbols).

\begin{figure}
\begin{center}
\resizebox{\hsize}{!}{\includegraphics{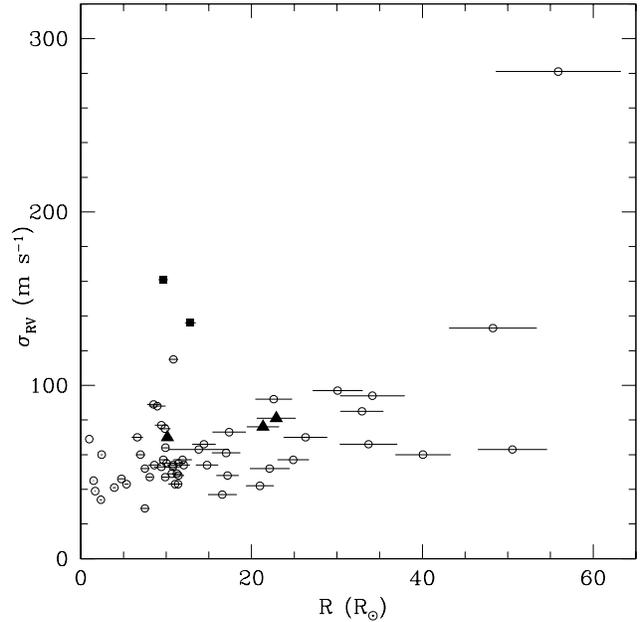}}
\end{center}
\caption{
        $\sigma_{\rm RV}$ versus stellar radius diagram for our sample, excluding 11 binaries.
        Refer to Fig.~\protect\ref{hrd} for the meaning of the symbols.
}
\label{rv_r}
\end{figure}

A similar trend is seen in the plot of $\sigma_{\rm RV}$ against stellar radius
in Fig.~\ref{rv_r}.
The two stars suspected to host brown dwarf companions, HD~27256 and HD~224533,
stand out as showing too large a $\sigma_{\rm RV}$ for their radii.
If these two stars are excluded, an unweighted least squares fit to the data
results in the relation
\[
\sigma_{\rm RV} ({\rm m\,s}^{-1}) = (1.76\pm0.31)\,R(R_\odot) + (38.1\pm6.0) \,.
\]
The Spearman rank correlation coefficient is 0.45.
This result, although supporting the increase of $\sigma_{\rm RV}$ with $R$,
is not significant because $\sigma_{\rm RV}$ is limited by the long-term
precision of our RV measurements
(about 25~${\rm m\,s}^{-1}$, see Setiawan et al. 2003, 2004).
 
\subsection{Distribution of metallicity }

Another major result of this study is the derivation of the age--metallicity relation
for the Solar Neighbourhood (see Fig.~\ref{amr}).
Its behaviour agrees, in general, with most previous determinations in the
literature, except for the very low \feh\ spread that we find for the youngest ages,
which is comparable to the observational error.
The main novelty with respect to previous results is that we derive the AMR
{\em using data for field red giants only}, whereas the majority of present-day
determinations have used samples of field dwarfs (including just a small
fraction of subgiants), or giants belonging to open clusters.
It worth noticing that we used {\em very simple data} in our determinations,
namely the $BV$ photometry together with Hipparcos parallaxes and measured
values for \Teff\ and \feh\ .

Of course, the same work can be extended to all giants in Hipparcos catalog,
once we have obtained homogeneous \Teff\ and \feh\ determinations for them.
This opens the possibility of improving considerably the statistics and
reliability of the local age--metallicity relation, simply by acquiring
spectroscopic data for a larger sample of bright giants, and performing
the same abundance and parameter analysis as in the present work.

Moreover, similar methods can be applied to nearby galaxies with well-known distances,
once we have available both the photometry and spectroscopy for a sufficiently large
number of their red giants.
Zaggia et al. (in preparation) use a procedure similar to ours to derive the
age--metallicity relation of the Sgr dSph galaxy.

\subsection{Stars hosting low mass companions and planets }

In the course of earlier studies we have identified three stars as candidates to
host low mass companions: two with planets (HD~47536 and HD~122430) and one with either
a brown dwarf or a planet companion (HD~11977).
We investigate a range of stellar masses larger than the range of masses usually investigated with radial velocity techniques. 
Not only can we provide better constraints on the companion mass, but we can
also investigate to what extent the conditions for companion formation differ
within the mass range.
Only one of our three stars, HD~122430, has nearly solar metallicity ($\feh=-0.05$),
while HD~11977 is slightly sub-solar ($\feh=-0.21$) and HD~47536 ($\feh=-0.68$)
is the most metal poor star of the sample.
Schuler et al. (2005) derived a metallicity of [Fe/H] = -0.58
for the K giant star hosting planet HD 13189. Although we are considering
a small number of objects, this result seems at odds with what has been
found for dwarf stars hosting giant exoplanets, which are preferentially
metal rich (e.g. Santos 2004). However, most RV planet search programs
have concentrated on solar mass stars and two of our planet hosting
giant stars have masses considerably larger than solar. In the case of
HD 13189 the host star has a mass of 3.5 M$_\odot$
(Schuler et al. 2005). At the present time
is is unknown what role stellar mass plays in the process of planet formation
and for massive stars this may be a more dominant factor than the metallicity.
Any investigation of the metallicity-planet relation among giant
stars should focus on those in the same mass range. It may be that
for a given mass range stars with higher metal abundances still tend to host 
a higher frequency of giant planets.

\begin{acknowledgements}
We are grateful to G. Caira for measuring most of the equivalent widths used here.
We thank A. Richichi and I. Percheron for their useful comments about stellar diameters, L. Portinari for discussions about the AMR,  and the anonymous referee for useful remarks.
The work by L.G. was funded by the MIUR COFIN 2004.
This project has benefitted from the support of the ESO DGDF.
L. da S. thanks the CNPq, Brazilian Agency, for the grants 30137/86-7 and 304134-2003.1. J. R. M. acknowledges continuous financial support of the CNPq and
FAPERN Brazilian Agencies.
\end{acknowledgements}

%

\begin{longtable}{l|rrr|rrrrrrrrrrrrrr}
\caption{
        \label{tabelao}
        Stellar parameters as derived from the observed \mv, \Teff, and \feh\ via the PDF method.
        Errors in \Teff\ and \feh\ are 70~K and 0.05~dex for all stars.
        The \feh\ values shown here were corrected by a zero-point offset of $-0.07$~dex.
        \ebv\ stands for the observed \bv\ minus the estimated intrinsic \bvo.
}
\\
\hline\hline
HD & \mv & \Teff & \feh & $t$ & $M$ &
\logg & $R$ & \ebv & $\theta$ \\
   &  (mag) & (K) & (dex)  & (Gyr) & (\Msun) &
(c.g.s.) & ($R_\odot$) & (mag) &  (mas) \\
\hline
\endfirsthead
\caption{continued.}\\
\hline\hline
HD & \mv & \Teff & \feh & $t$ & $M$ &
\logg & $R$ & \ebv & $\theta$ \\
   &  (mag) & (K) & (dex)  & (Gyr) & (\Msun) &
(c.g.s.) & ($R_\odot$) & (mag) &  (mas) \\
\hline
\endhead
\hline
\endfoot
2114 & $-0.53\pm0.41$ & $5288$ & $-0.03$ & $0.32\pm0.07$ & $3.16\pm0.29$ & $2.62\pm0.18$ & $13.8\pm3.5$ & $-0.00\pm0.04$ & $0.71\pm0.31$ \\ 
2151 & $3.45\pm0.01$ & $5964$ & $-0.03$ & $5.12\pm1.03$ & $1.17\pm0.05$ & $4.02\pm0.04$ & $1.69\pm0.05$ & $0.01\pm0.02$ & $2.10\pm0.07$ \\ 
7672 & $0.95\pm0.12$ & $5096$ & $-0.33$ & $1.07\pm0.29$ & $1.95\pm0.20$ & $2.87\pm0.07$ & $8.18\pm0.46$ & $0.01\pm0.02$ & $0.97\pm0.11$ \\ 
11977 & $0.57\pm0.07$ & $4975$ & $-0.21$ & $1.30\pm0.48$ & $1.87\pm0.30$ & $2.66\pm0.10$ & $10.2\pm0.5$ & $-0.02\pm0.05$ & $1.42\pm0.11$ \\ 
12438 & $0.68\pm0.14$ & $4975$ & $-0.61$ & $5.52\pm2.77$ & $1.02\pm0.19$ & $2.44\pm0.07$ & $9.67\pm0.48$ & $-0.00\pm0.02$ & $1.05\pm0.12$ \\ 
16417 & $3.74\pm0.04$ & $5936$ & $0.19$ & $4.34\pm0.82$ & $1.18\pm0.04$ & $4.12\pm0.03$ & $1.50\pm0.05$ & $0.00\pm0.02$ & $0.55\pm0.03$ \\ 
18322 & $0.83\pm0.06$ & $4637$ & $-0.07$ & $4.18\pm1.91$ & $1.21\pm0.20$ & $2.42\pm0.08$ & $10.8\pm0.5$ & $-0.02\pm0.04$ & $2.46\pm0.18$ \\ 
18885 & $1.15\pm0.14$ & $4737$ & $0.10$ & $1.64\pm0.91$ & $1.76\pm0.38$ & $2.70\pm0.16$ & $9.44\pm0.86$ & $-0.00\pm0.05$ & $1.01\pm0.16$ \\ 
18907 & $3.47\pm0.05$ & $5091$ & $-0.61$ & $10.69\pm0.74$ & $0.93\pm0.03$ & $3.59\pm0.03$ & $2.45\pm0.09$ & $-0.02\pm0.02$ & $0.75\pm0.04$ \\ 
21120 & $-0.45\pm0.18$ & $5180$ & $-0.12$ & $0.38\pm0.05$ & $3.01\pm0.14$ & $2.52\pm0.06$ & $15.1\pm1.2$ & $-0.03\pm0.03$ & $2.17\pm0.36$ \\ 
22663 & $0.43\pm0.08$ & $4624$ & $0.11$ & $1.34\pm0.53$ & $1.94\pm0.30$ & $2.44\pm0.11$ & $13.3\pm0.9$ & $-0.13\pm0.05$ & $1.85\pm0.19$ \\ 
23319 & $0.91\pm0.07$ & $4522$ & $0.24$ & $3.12\pm1.15$ & $1.38\pm0.19$ & $2.43\pm0.09$ & $11.4\pm0.6$ & $-0.03\pm0.03$ & $1.95\pm0.15$ \\ 
23940 & $0.83\pm0.12$ & $4884$ & $-0.35$ & $5.81\pm3.17$ & $1.01\pm0.21$ & $2.42\pm0.08$ & $9.90\pm0.44$ & $-0.01\pm0.02$ & $1.06\pm0.11$ \\ 
26923 & $4.69\pm0.05$ & $6207$ & $-0.06$ & $0.66\pm0.58$ & $1.06\pm0.02$ & $4.45\pm0.02$ & $1.00\pm0.00$ & $-0.01\pm0.01$ & $0.44\pm0.01$ \\ 
27256 & $-0.17\pm0.05$ & $5196$ & $0.07$ & $0.33\pm0.02$ & $3.11\pm0.06$ & $2.69\pm0.04$ & $12.8\pm0.6$ & $0.01\pm0.03$ & $2.39\pm0.17$ \\ 
27371 & $0.28\pm0.12$ & $5030$ & $0.13$ & $0.53\pm0.09$ & $2.70\pm0.13$ & $2.67\pm0.04$ & $12.1\pm0.7$ & $-0.05\pm0.02$ & $2.38\pm0.28$ \\ 
27697 & $0.41\pm0.10$ & $4951$ & $0.06$ & $0.64\pm0.13$ & $2.54\pm0.14$ & $2.70\pm0.04$ & $11.3\pm0.6$ & $-0.03\pm0.03$ & $2.25\pm0.21$ \\ 
32887 & $-1.02\pm0.10$ & $4131$ & $-0.09$ & $1.72\pm0.47$ & $1.70\pm0.19$ & $1.43\pm0.09$ & $40.1\pm3.2$ & $0.03\pm0.06$ & $5.37\pm0.69$ \\ 
34642 & $2.17\pm0.04$ & $4870$ & $-0.04$ & $3.40\pm1.09$ & $1.38\pm0.12$ & $3.09\pm0.07$ & $5.36\pm0.24$ & $-0.03\pm0.04$ & $1.48\pm0.09$ \\ 
36189 & $-0.64\pm0.16$ & $5081$ & $-0.02$ & $0.32\pm0.04$ & $3.23\pm0.13$ & $2.44\pm0.05$ & $17.2\pm1.3$ & $0.02\pm0.03$ & $1.12\pm0.17$ \\ 
36848 & $1.83\pm0.06$ & $4460$ & $0.21$ & $7.58\pm2.18$ & $1.13\pm0.09$ & $2.64\pm0.06$ & $8.08\pm0.36$ & $-0.01\pm0.02$ & $1.42\pm0.10$ \\ 
47205 & $2.46\pm0.03$ & $4744$ & $0.18$ & $4.17\pm1.32$ & $1.32\pm0.12$ & $3.11\pm0.07$ & $5.08\pm0.23$ & $-0.07\pm0.05$ & $2.38\pm0.14$ \\ 
47536 & $-0.17\pm0.15$ & $4352$ & $-0.68$ & $9.33\pm1.88$ & $0.94\pm0.06$ & $1.72\pm0.08$ & $21.3\pm1.9$ & $0.01\pm0.03$ & $1.64\pm0.25$ \\ 
50778 & $-0.36\pm0.15$ & $4084$ & $-0.29$ & $8.11\pm2.37$ & $1.03\pm0.10$ & $1.46\pm0.10$ & $30.1\pm2.9$ & $0.03\pm0.05$ & $3.62\pm0.60$ \\ 
61935 & $0.71\pm0.08$ & $4879$ & $-0.01$ & $1.18\pm0.42$ & $2.02\pm0.29$ & $2.71\pm0.09$ & $10.1\pm0.5$ & $0.00\pm0.04$ & $2.11\pm0.18$ \\ 
62644 & $3.13\pm0.04$ & $5526$ & $0.12$ & $3.78\pm0.19$ & $1.31\pm0.03$ & $3.78\pm0.03$ & $2.35\pm0.09$ & $-0.00\pm0.03$ & $0.90\pm0.05$ \\ 
62902 & $1.17\pm0.15$ & $4311$ & $0.33$ & $6.95\pm2.57$ & $1.10\pm0.16$ & $2.29\pm0.09$ & $11.9\pm1.1$ & $0.06\pm0.04$ & $1.52\pm0.24$ \\ 
63697 & $0.72\pm0.12$ & $4322$ & $0.13$ & $4.62\pm2.17$ & $1.26\pm0.21$ & $2.16\pm0.10$ & $14.8\pm1.3$ & $-0.02\pm0.04$ & $1.77\pm0.26$ \\ 
65695 & $0.51\pm0.16$ & $4468$ & $-0.14$ & $4.29\pm2.17$ & $1.27\pm0.20$ & $2.19\pm0.10$ & $14.4\pm1.4$ & $0.01\pm0.04$ & $1.75\pm0.30$ \\ 
70982 & $0.34\pm0.15$ & $5089$ & $-0.03$ & $0.61\pm0.12$ & $2.54\pm0.15$ & $2.72\pm0.05$ & $11.0\pm0.8$ & $-0.03\pm0.02$ & $0.72\pm0.10$ \\ 
72650 & $0.45\pm0.15$ & $4310$ & $0.06$ & $4.05\pm1.94$ & $1.32\pm0.20$ & $2.06\pm0.11$ & $17.0\pm1.6$ & $-0.01\pm0.04$ & $1.05\pm0.18$ \\ 
76376 & $0.17\pm0.20$ & $4282$ & $-0.10$ & $5.39\pm2.56$ & $1.20\pm0.17$ & $1.87\pm0.13$ & $20.1\pm2.4$ & $0.02\pm0.04$ & $1.44\pm0.30$ \\ 
81797 & $-1.69\pm0.09$ & $4186$ & $0.00$ & $0.42\pm0.16$ & $3.03\pm0.36$ & $1.48\pm0.10$ & $50.5\pm4.0$ & $0.04\pm0.04$ & $8.65\pm1.06$ \\ 
83441 & $0.89\pm0.15$ & $4649$ & $0.10$ & $2.98\pm1.57$ & $1.40\pm0.28$ & $2.48\pm0.11$ & $10.8\pm0.6$ & $-0.03\pm0.03$ & $0.98\pm0.12$ \\ 
85035 & $2.62\pm0.14$ & $4680$ & $0.12$ & $6.42\pm2.22$ & $1.17\pm0.11$ & $3.12\pm0.08$ & $4.77\pm0.37$ & $-0.03\pm0.03$ & $0.58\pm0.08$ \\ 
90957 & $0.21\pm0.17$ & $4172$ & $0.05$ & $5.57\pm2.46$ & $1.18\pm0.18$ & $1.79\pm0.13$ & $22.1\pm2.3$ & $0.02\pm0.04$ & $1.73\pm0.32$ \\ 
92588 & $3.57\pm0.06$ & $5136$ & $0.07$ & $4.72\pm0.53$ & $1.24\pm0.04$ & $3.75\pm0.03$ & $2.36\pm0.12$ & $-0.03\pm0.03$ & $0.64\pm0.05$ \\ 
93257 & $1.81\pm0.09$ & $4607$ & $0.13$ & $4.95\pm2.09$ & $1.26\pm0.15$ & $2.75\pm0.09$ & $7.51\pm0.51$ & $-0.03\pm0.03$ & $1.28\pm0.14$ \\ 
93773 & $0.72\pm0.21$ & $4985$ & $-0.07$ & $0.96\pm0.27$ & $2.15\pm0.21$ & $2.79\pm0.07$ & $9.44\pm0.76$ & $-0.03\pm0.02$ & $0.61\pm0.11$ \\ 
99167 & $-0.43\pm0.18$ & $4010$ & $-0.36$ & $8.92\pm2.10$ & $0.99\pm0.08$ & $1.33\pm0.10$ & $34.1\pm3.8$ & $0.14\pm0.04$ & $2.84\pm0.55$ \\ 
101321 & $1.72\pm0.19$ & $4803$ & $-0.14$ & $3.96\pm1.62$ & $1.31\pm0.14$ & $2.88\pm0.10$ & $6.62\pm0.66$ & $-0.05\pm0.03$ & $0.59\pm0.11$ \\ 
107446 & $-0.63\pm0.09$ & $4148$ & $-0.10$ & $2.83\pm1.20$ & $1.42\pm0.22$ & $1.52\pm0.11$ & $32.9\pm2.5$ & $-0.02\pm0.07$ & $4.38\pm0.51$ \\ 
110014 & $-0.29\pm0.19$ & $4445$ & $0.19$ & $0.86\pm0.34$ & $2.30\pm0.28$ & $2.06\pm0.10$ & $22.6\pm2.2$ & $-0.05\pm0.04$ & $2.15\pm0.39$ \\ 
111884 & $0.58\pm0.17$ & $4271$ & $-0.06$ & $7.27\pm2.62$ & $1.12\pm0.11$ & $2.02\pm0.10$ & $16.6\pm1.7$ & $0.01\pm0.04$ & $1.33\pm0.24$ \\ 
113226 & $0.37\pm0.06$ & $5086$ & $0.09$ & $0.56\pm0.09$ & $2.64\pm0.11$ & $2.71\pm0.04$ & $11.4\pm0.5$ & $-0.07\pm0.03$ & $3.38\pm0.23$ \\ 
115478 & $0.53\pm0.20$ & $4250$ & $0.03$ & $6.05\pm2.54$ & $1.18\pm0.14$ & $2.00\pm0.11$ & $17.4\pm2.0$ & $-0.03\pm0.04$ & $1.77\pm0.36$ \\ 
122430 & $-0.15\pm0.21$ & $4300$ & $-0.05$ & $3.11\pm1.70$ & $1.39\pm0.27$ & $1.83\pm0.12$ & $22.9\pm2.3$ & $0.01\pm0.04$ & $1.60\pm0.32$ \\ 
124882 & $-0.35\pm0.09$ & $4293$ & $-0.24$ & $4.30\pm2.07$ & $1.21\pm0.21$ & $1.69\pm0.11$ & $24.9\pm1.8$ & $0.01\pm0.30$ & $2.71\pm0.31$ \\ 
125560 & $1.01\pm0.09$ & $4472$ & $0.16$ & $5.21\pm2.28$ & $1.14\pm0.19$ & $2.36\pm0.08$ & $11.3\pm0.5$ & $0.01\pm0.03$ & $1.79\pm0.15$ \\ 
131109 & $-0.22\pm0.16$ & $4158$ & $-0.07$ & $5.14\pm2.43$ & $1.18\pm0.20$ & $1.64\pm0.11$ & $26.3\pm2.6$ & $0.05\pm0.04$ & $1.87\pm0.32$ \\ 
136014 & $0.83\pm0.21$ & $4869$ & $-0.46$ & $5.32\pm2.80$ & $1.05\pm0.21$ & $2.44\pm0.09$ & $9.86\pm0.66$ & $0.01\pm0.03$ & $0.78\pm0.13$ \\ 
148760 & $1.87\pm0.12$ & $4654$ & $0.13$ & $4.16\pm1.61$ & $1.32\pm0.15$ & $2.84\pm0.09$ & $6.99\pm0.54$ & $-0.05\pm0.03$ & $0.94\pm0.13$ \\ 
151249 & $-1.14\pm0.14$ & $3886$ & $-0.37$ & $7.11\pm2.34$ & $1.02\pm0.12$ & $0.92\pm0.12$ & $55.9\pm7.3$ & $0.02\pm0.04$ & $5.42\pm1.06$ \\ 
152334 & $0.30\pm0.09$ & $4169$ & $0.06$ & $5.80\pm2.26$ & $1.19\pm0.14$ & $1.84\pm0.10$ & $21.0\pm1.6$ & $0.00\pm0.04$ & $4.23\pm0.49$ \\ 
152980 & $-0.79\pm0.16$ & $4176$ & $0.01$ & $1.70\pm0.57$ & $1.74\pm0.24$ & $1.59\pm0.11$ & $33.7\pm3.4$ & $0.04\pm0.04$ & $3.36\pm0.58$ \\ 
159194 & $1.61\pm0.20$ & $4444$ & $0.14$ & $6.76\pm2.42$ & $1.15\pm0.13$ & $2.56\pm0.11$ & $8.97\pm0.98$ & $0.02\pm0.03$ & $0.78\pm0.16$ \\ 
165760 & $0.33\pm0.13$ & $5005$ & $0.02$ & $0.60\pm0.12$ & $2.59\pm0.15$ & $2.70\pm0.05$ & $11.5\pm0.7$ & $-0.04\pm0.04$ & $1.46\pm0.18$ \\ 
169370 & $1.61\pm0.18$ & $4460$ & $-0.17$ & $9.31\pm1.90$ & $1.05\pm0.06$ & $2.56\pm0.07$ & $8.61\pm0.70$ & $0.03\pm0.02$ & $0.93\pm0.15$ \\ 
174295 & $0.41\pm0.14$ & $4893$ & $-0.24$ & $1.87\pm0.86$ & $1.60\pm0.31$ & $2.54\pm0.10$ & $10.8\pm0.7$ & $-0.02\pm0.03$ & $1.12\pm0.14$ \\ 
175751 & $0.82\pm0.12$ & $4710$ & $0.01$ & $2.78\pm1.39$ & $1.42\pm0.27$ & $2.50\pm0.10$ & $10.7\pm0.5$ & $-0.05\pm0.06$ & $1.56\pm0.16$ \\ 
177389 & $2.48\pm0.05$ & $5131$ & $0.02$ & $1.87\pm0.15$ & $1.65\pm0.05$ & $3.43\pm0.04$ & $3.93\pm0.19$ & $-0.01\pm0.03$ & $0.99\pm0.07$ \\ 
179799 & $2.35\pm0.17$ & $4865$ & $0.03$ & $3.61\pm1.21$ & $1.36\pm0.13$ & $3.17\pm0.09$ & $4.80\pm0.43$ & $-0.05\pm0.03$ & $0.65\pm0.11$ \\ 
187195 & $1.29\pm0.14$ & $4444$ & $0.13$ & $6.44\pm2.81$ & $1.09\pm0.19$ & $2.39\pm0.11$ & $10.6\pm0.7$ & $0.01\pm0.03$ & $1.13\pm0.15$ \\ 
189319 & $-1.11\pm0.13$ & $3978$ & $-0.29$ & $4.79\pm2.03$ & $1.17\pm0.17$ & $1.10\pm0.11$ & $48.3\pm5.1$ & $0.07\pm0.04$ & $5.34\pm0.89$ \\ 
190608 & $1.61\pm0.08$ & $4741$ & $0.05$ & $2.77\pm0.94$ & $1.48\pm0.15$ & $2.82\pm0.08$ & $7.51\pm0.44$ & $-0.04\pm0.03$ & $1.41\pm0.13$ \\ 
198232 & $-0.45\pm0.20$ & $4923$ & $0.03$ & $0.36\pm0.07$ & $3.13\pm0.18$ & $2.47\pm0.07$ & $16.4\pm1.6$ & $-0.02\pm0.03$ & $1.30\pm0.25$ \\ 
198431 & $1.45\pm0.17$ & $4641$ & $-0.12$ & $6.08\pm2.53$ & $1.17\pm0.14$ & $2.61\pm0.10$ & $8.53\pm0.76$ & $-0.03\pm0.03$ & $1.04\pm0.17$ \\ 
199665 & $1.19\pm0.11$ & $5089$ & $0.05$ & $0.94\pm0.18$ & $2.13\pm0.13$ & $2.93\pm0.08$ & $8.00\pm0.63$ & $-0.05\pm0.03$ & $1.02\pm0.13$ \\ 
217428 & $-0.24\pm0.33$ & $5285$ & $0.03$ & $0.36\pm0.08$ & $2.99\pm0.21$ & $2.75\pm0.14$ & $11.6\pm1.9$ & $0.04\pm0.03$ & $0.55\pm0.17$ \\ 
218527 & $0.75\pm0.24$ & $5084$ & $0.03$ & $0.79\pm0.19$ & $2.32\pm0.18$ & $2.81\pm0.06$ & $9.57\pm0.85$ & $-0.08\pm0.02$ & $1.04\pm0.21$ \\ 
219615 & $0.68\pm0.08$ & $4885$ & $-0.51$ & $5.46\pm2.50$ & $1.03\pm0.17$ & $2.43\pm0.06$ & $9.92\pm0.42$ & $-0.01\pm0.03$ & $2.30\pm0.18$ \\ 
224533 & $0.70\pm0.12$ & $5062$ & $0.00$ & $0.80\pm0.18$ & $2.31\pm0.16$ & $2.80\pm0.04$ & $9.65\pm0.54$ & $-0.05\pm0.02$ & $1.31\pm0.15$ \\

\end{longtable}

\end{document}